\documentclass[journal]{IEEEtran}

\usepackage[utf8]{inputenc} 
\usepackage[T1]{fontenc}    
\usepackage{hyperref}       
\usepackage{url}            
\usepackage{booktabs}       
\usepackage{amsfonts}       
\usepackage{amsmath}
\usepackage{microtype}      
\usepackage{graphicx}       
\graphicspath{{media/}}     
\usepackage[shortlabels]{enumitem}
\usepackage{enumerate}
\usepackage{booktabs,array}
\usepackage{pgfplots} 
\usepackage{pgfplotstable} 
\usepgfplotslibrary{groupplots} 
\pgfplotsset{compat=newest}
\usepackage{tikz}
\usetikzlibrary{shapes.geometric, arrows, calc, fit,positioning}
\usepackage{pgfplots}
\usepackage{subfig}
\usepackage{caption}
\usetikzlibrary{arrows.meta}
\usepackage{multirow}
\usepackage{multicol}
\usepackage{makecell}

\usepgfplotslibrary{statistics}
\usetikzlibrary{pgfplots.statistics}
\usepackage{booktabs}

\usepackage{cite}

\usepackage{balance}

\usepackage{authblk}
\newtheorem{remark}{Remark}

\pgfplotsset{compat = newest}
\usepackage{mathtools}
\usepackage{comment}
\makeatletter
\newcommand{\vast}{\bBigg@{4}}
\newcommand{\Vast}{\bBigg@{5}}
\DeclareMathOperator*{\argmax}{argmax}

\makeatother

\begin{document}

\title{Centrality-Based Deployment of Queue Policies\\ in Acyclic Multipath Routing Networks}

\author{Mahima Gupta, Acquin Biju,
        Rijul Jain\IEEEauthorrefmark{1}, Dipesh Sharma\IEEEauthorrefmark{1}
        and~Sreelakshmi Manjunath\IEEEauthorrefmark{2}
\thanks{The authors are with the School
of Computing and Electrical Engineering, Indian Institute of Technology Mandi,
HP, 170075 India. (e-mail: \{t22055, s24122, b20126, b20096\}@students.iitmandi.ac.in, sreelakshmi@iitmandi.ac.in). The work is supported by the Science and Engineering Research Board, Department of Science and Technology, Government of India through grant SRG/2021/002269.}
\thanks{\IEEEauthorrefmark{1}These authors have made equal contribution. \IEEEauthorrefmark{2}Corresponding author.}
}

\maketitle

\begin{abstract}
Excessive queueing delays constitute a significant impediment to latency-sensitive network applications. Although effective deployment of Active Queue Management (AQM) strategies has been proposed as a necessary solution, deployment remains sparse. This paper studies AQM deployment in a specific class of networks where routers/switches have a topological hierarchy, form acyclic paths, and adopt multipath routing. Our approach rests on the well-established premise that AQM deployment impacts packet-forwarding dynamics in networks carrying TCP flows, thus establishing a direct link between stability and network performance. We use fluid models for TCP and queue dynamics in the network, along with a simple threshold-based queue policy to outline a closed-loop model for the network. Stability analyses reveal that while the network is vulnerable to instability as the average round-trip time (RTT) of the TCP flows increases, it tolerates a much larger RTT without losing stability when the threshold-based AQM is deployed in an appropriate router. We then define a Katz centrality-based metric to choose the most appropriate router for AQM deployment, and argue that doing so ensures the greatest stabilising effect. Finally, packet-level simulations corroborate that the proposed deployment strategy ensures low-latency operation of the network.
\end{abstract}

\begin{IEEEkeywords}
Active queue management, bufferbloat, latency, performance, centrality measures.
\end{IEEEkeywords}

\section{Introduction}
Increased latency can significantly impair the performance of interactive and transaction-based applications in today's Internet. It is primarily attributed to large router buffers that are persistently full\textemdash a phenomenon known as \emph{bufferbloat}~\cite{gettys2012bufferbloat,allman2012comments,cerf2014bufferbloat,McFillin2022}. Enhanced transport-layer protocols are commonly viewed as a primary solution to mitigate bufferbloat~\cite{cardwell2017bbr,ferlin2014tackling}. Transmission Control Protocol (TCP), the primary transport-layer protocol, incorporates congestion-avoidance algorithms, which force end hosts to reduce their sending rate upon receiving congestion signals such as packet loss, increased round-trip time (RTT), or a combination of both. 
 Although congestion-avoidance algorithms are necessary, they are insufficient; network devices must use appropriate mechanisms that can complement the action of end hosts by providing appropriate congestion feedback~\cite{baker2015ietf}. 
 
 The use of effective Active Queue Management (AQM) strategies, which can manage queues by dropping or marking packets appropriately, in packet-forwarding devices has been identified as a necessary solution to bufferbloat~\cite{gettys2012bufferbloat,nichols2012controlling, harrison2023buffer}. Numerous AQM policies have been proposed; for  surveys, see~\cite{alwahab2018simulation,toopchinezhad2025machine}. Most Linux-based routers have some version of the Random Early Detection (RED)~\cite{floyd1993random} queue policy implemented in them, and Proportional Integral Enhanced (PIE) is implemented in some broadband access devices~\cite{pan2013pie}. However, due to the need for precise tuning of AQM parameters, as well as the scale and heterogeneity of the modern-day Internet, AQM deployment remains sparse and most devices continue to use the simple DropTail policy~\cite{bideh2016tada,de2023rfc}. 

Effective AQM deployment raises two central questions: (i) which AQM policy should be selected from the many alternatives; and (ii) which routers should deploy it, so that network-wide deployment is unnecessary? Various studies on AQM policies have addressed the first question. However, the second one is not so well studied. The relevance of this question is due to two reasons. First, indiscriminate network-wide deployment is rendered nearly-impossible due to heterogeneity. Secondly, such deployment could lead to unnecessary congestion signals thereby affecting goodput. Therefore, it is important to identify bottleneck links with congestion, and prioritize AQM deployment in network devices feeding these links. 

An early study in this direction establishes that deploying an AQM on a single router, in a given network, could have a positive impact on other non-AQM routers~\cite{mrozowski2009aqm}. However, it does not provide any guidelines on the choice of router. Yet another study~\cite{zhu2006edge} argued against network-wide deployment and proposed a scheme called Edge-based AQM\textemdash an architectural placement proposal that suggests that AQMs be deployed in edge routers alone. This idea of edge-based deployment is supported by IETF's recent RFC 9330~\cite{briscoe2023low}, although in the context of a specific AQM called L4S. Based on these recommendations, an AQM policy must be deployed in all edge routers, regardless of their number. Further, these recommendations disregard the possibility of a bottleneck link occurring elsewhere in the network, including the core. 
Given that AQMs provide active feedback to TCP hosts, the choice of router for AQM deployment in a network impacts the traffic flow in the entire network, and thereby impacts the queue sizes and delays at other non-AQM routers. Indeed, the IETF Best Current Practices for AQM (RFC 7567~\cite{baker2015ietf}), argues that lack of attention to the dynamics of packet forwarding can result in severe service degradation at the user end. Therefore, one must study the interaction of TCP and AQM in order to identify the best router for AQM deployment. In addition, a solution emerging from such a study may need a global view of the network, which could be challenging, in general. 

In this work, we consider a specific class of networks, where the packet-forwarding devices have a topological ordering, packets to a destination could be routed through multiple acyclic paths and a global view of the network is available. This is reflective of enterprise networks and data centre networks, where switches have a hierarchy and multipath routing algorithms are adopted. Further, these networks are often implemented over the SDN architecture, where routing is performed by an SDN controller that has a global view of the data-plane switches. Our work leverages this network structure to identify a router/switch at which an AQM must be deployed.

Our approach uses a control-theoretic perspective, which is well suited to gain insight into the interaction of TCP and AQM~\cite{srikant2004mathematics}. We use appropriate fluid models to represent the congestion-window evolution of TCP hosts, and the queue dynamics in a network of routers/switches to understand the closed-loop dynamics of the network. The traffic flow within the network, dictated by a routing algorithm, is assumed to be known in the form of a flow-dynamics matrix that feeds into the queue-dynamics model. We consider a simple threshold-based queue policy which is deployed at exactly one router/switch in the network. It is well known that stability and performance are strongly coupled in TCP-AQM networks~\cite{srikant2004mathematics,hollot2002analysis,raina2005part}. Therefore, we conduct a detailed stability analysis of the closed-loop network model. The stability results enable us to identify the \emph{target} router/switch which, when deployed with the AQM, would have the most stabilising effect on the closed-loop dynamics. We then establish that the target router/switch can be identified by using Katz centrality of the graph underlying the traffic flow in the network. Through our analysis, model-based computations and packet-level simulations, we establish that deploying the AQM at a packet-forwarding device that has the largest Katz centrality leads to improved latency in the network. Given that an SDN controller has a global view of the traffic flow in the network, we argue that the required Katz centrality can be computed through a simple matrix equation, ensuring scalability of the proposed solution.

The remainder of the paper is organized as follows. Section~\ref{sec:MathematicalModelling} outlines the models. In Section~\ref{sec:Stability_Analysis}, we derive a sufficient condition, as well as the necessary and sufficient condition, for local stability of the network. We present the guidelines AQM deployment in Section~\ref{sec:Centrality_Guidelines}. In Section~\ref{sec:Simulations}, we validate the design guidelines through a simulation-based performance evaluation. The conclusions are presented in Section~\ref{sec:Conclude}. 
\section{Mathematical modelling}
\label{sec:MathematicalModelling}
Consider a network with one edge router and $N_R$ routers in the core network. There are $N_T$ end hosts feeding into the edge router. We assume that the end systems are the sources of long-lived TCP flows and use TCP variants which use the Additive Increase Multiplicative Decrease (AIMD) based congestion avoidance mechanism. We first outline the fluid model that governs the evolution of the TCP congestion window and then describe the models for the evolution of the queue size at the routers. This is followed by a description of the Active Queue Management (AQM) policy to be deployed and the closed-loop model for the overall network with the AQM. 

\subsection{Transmission control protocol: fluid model}
Transmission control protocol (TCP) uses a sliding window mechanism, that enables it to send a set of packets sequentially, instead of waiting for each packet to be acknowledged before sending the next. The size of this sending window is increased or reduced based on the congestion feedback received from the network. There are different variants of TCP, and these variants differ mainly in the form of feedback they use to infer congestion.

Some flavours of TCP use either packet loss or an estimate of queueing delay as congestion feedback, while others use a combination of the two. Tahoe TCP and TCP Reno are two of the earliest proposals for loss-based TCP.  
While, Vegas TCP, arguably, laid the foundations for delay-based TCP flavours. Recognising that some issues such as efficiency, RTT fairness and TCP fairness can not be simultaneously mitigated either by loss- or delay-based protocols, the authors of~\cite{tan2006compound} proposed Compound TCP, which is a synergy of the two. It is currently implemented in the Windows OS. Other variants of loss- and delay-based protocols are TCP Illinois and TCP-Africa.

We recapitulate the development of the fluid model for TCP presented in~\cite{misra2000fluid,raina2005buffer}. Following this, we present a general fluid model for the window dynamics of a class of delay- and loss-based protocols. Let $W(t)$ represent the size of the sending window (number of packets sent in one round-trip time) of a \emph{single} TCP flow of round-trip time $\tau$, which is increased by $1$ packet per RTT and reduced to $W/2$ every time a packet loss is detected. If the rate at which packets are sent is approximated as $W(t)/\tau$, then acknowledgements are received at a rate of $W(t-\tau)/\tau$ at time $t$. Let $p(t)$ be the packet-drop probability at time $t$, dictated by an AQM in the network. Suppose that there are $N$ end hosts, and let $W^N(t)$ represent the \emph{sum} of all the window sizes. Then, $W^N(t)$ follows
\begin{align*}
 W^N(t+\delta)-W^N(t) \approx \frac{\delta \, N}{\tau} - \delta \frac{W^N(t)}{2N} \bigg(\frac{W^N(t-\tau)}{\tau}p(t-\tau)\bigg).
\end{align*}
Upon dividing both sides of the above equation by $N$, we arrive at the following continuous time approximation for the update of the \emph{average} window size $w(t)=W^N(t)/N$
\begin{align*}
 \dot{w}(t) = \frac{1}{\tau}-\frac{w(t)}{2}\bigg(\frac{w(t-\tau)}{\tau}p(t-\tau)\bigg),
\end{align*}
where $\dot{w}(t) = dw(t)/dt$ and $\tau$ can be regarded as the average RTT of the TCP flows. 

Now consider a TCP variant that increases the sending window by $i(w)$ per acknowledgement received and reduces it by $d(w)$ per packet drop. If a packet is lost with a probability of $p(t)$, it is acknowledged with a probability $1-p(t)$. The average window size $w(t)$ is updated as~\cite{raina2005part}
\begin{align}
 \dot{w}(t) = \bigg(i\big(w(t)\big)\big(1-p(t-\tau)\big)-d\big(w(t)\big)p(t-\tau)\bigg)\frac{w(t-\tau)}{\tau}.\label{eq:gen_TCP}
\end{align}
Note that equation~\eqref{eq:gen_TCP} is a nonlinear delay-differential equation, where the RTT $\tau$ acts as the feedback delay. Equation~\eqref{eq:gen_TCP} serves as a generalised fluid model for multiple TCP variants which follow the Additive Increase Multiplicative Decrease (AIMD) algorithm for congestion avoidance. 
The functions $i(w)$ and $d(w)$ are specific to each TCP variant. 
The functional forms $i(w)$ and $d(w)$ for some delay- and loss-based TCP variants are listed in Table~\ref{tab:TCP_functional_forms}.

\begin{table}[t]
\centering
\caption{Functional forms for the window update functions of some TCP variants based on Additive Increase (AI) Multiplicative Decrease (MD) congestion-control.}
\begin{tabular}{l | l | l | l}
\hline\label{tab:TCP_functional_forms}
\rule{0pt}{1.5\normalbaselineskip}\hspace{-1mm} TCP variant & AI: $i(w)$ & MD: $d(w)$ & Default \\ [2ex]
\hline\rule{0pt}{1.5\normalbaselineskip}
\hspace{-1mm}
\multirow{3}{*}{Compound TCP~\cite{tan2006compound}} & \multirow{3}{*}{$\alpha\, w^{k-1}$} & \multirow{3}{*}{$\beta\,w$} & $\alpha = 0.125$\\
\hspace{-1mm}
& & & $k=0.75$\\
\hspace{-1mm}
& & & $\beta=0.5$\\[2ex]
\hline\rule{0pt}{1.5\normalbaselineskip}
\hspace{-1mm}
\multirow{2}{*}{TCP Illinois~\cite{liu2008tcp}} & \multirow{2}{*}{$\bar{\alpha}/w$} & \multirow{3}{*}{$\underline{\beta}\,w$} & $\bar\alpha = 10$\\
\hspace{-1mm}
& & & $\underline\beta = 0.125$\\[2ex]
\hline\rule{0pt}{1.5\normalbaselineskip}
\hspace{-1mm}
TCP Linux/NewReno~\cite{floyd2004rfc3782} & $1/w$ & $w/2$ & - \\
[2ex]
\hline\rule{0pt}{1.5\normalbaselineskip}
\hspace{-1mm}
Data Centre TCP~\cite{bensley2017data} & $1/ w$ & $(\bar{\alpha}/2)w$ & $\bar\alpha = 1$\\
\hline 
\end{tabular}
\end{table} 

\begin{remark}
Ideally, the RTT comprises the queueing delay and the propagation delay. Of these, the queueing delay depends on the queue sizes in the routers that a flow encounters, and varies with respect to time, while the propagation delay is constant. If this is to be captured in equation~\eqref{eq:gen_TCP}, we would have a delay-differential equation with time varying delay. Such models could generally be analytically intractable and may not lead to any analytical insight. We therefore assume a constant RTT, which yields the model in equation~\eqref{eq:gen_TCP} that lends itself to control-theoretic analysis.   
\end{remark}

\subsection{Network of routers: model for queue evolution}
\label{sec:QueueEvolution}
We first outline a model for the evolution of the queue size. For a set of $N_T$ TCP sources, with aggregate window size $W^{N}(t)$, let the $X^N(t) = W^N(t)/\tau$ be an approximation of the rate at which packets arrive at the queue at time $t$, and let the average arrival be $x(t) = X^N(t)/N_T$. Likewise, the average arrival rate can also be written in terms of the average window size as $x(t) = w(t)/\tau$. If packets are dropped with a probability $p(\cdot)$, then they would be queued with a probability $1-p(\cdot)$. Thus, the total arrival at the queue can be written as $\big(1-p(t)\big)\delta N_Tx(t)$. Next, if $C$ is the service capacity of the outgoing link of the router, in the time interval $(t,t+\delta)$, the queue size $Q^N(t)$ evolves according to~\cite{raina2005buffer}
\begin{align}
 Q^N(t+\delta) \approx \Big[Q^N(t) + \big(1-p(t)\big)\delta N_Tx(t) - \delta C\Big]^B_0,\label{eq:total_queue_evolution}
\end{align}
where the notation $[q]^b_0=\min\big(\max(q,0),b\big)$ is used, and $B$ represents the size of the router buffer.

Now, consider a regime where the buffer is sized as per the bandwidth-delay product rule.  Then, the queue size $q(t)$ evolution follows
\begin{align}
 \dot{\mathfrak{q}}(t) = \vast\{\begin{array}{l r}
             \big(1-p(t)\big)N_T\,\frac{w(t)}{\tau} - C & \text{ if } q(t) > 0,\\
             \\
             \text{max}\big\{ \big(1-p(t)\big)N_T\frac{w(t)}{\tau} - C,0\big\}& \text{ if } q(t) = 0,
              \end{array}\label{eq:single_queue_actualmodel}
\end{align}
which can be approximated using a continuous and differentiable activation function as 
\begin{align}
\dot{\mathfrak{q}}(t) = \Big(\big(1-p(t)\big)N_T\frac{w(t)}{\tau} - C\Big)\frac{1}{1+e^{-K\mathfrak{q}(t)}},\label{eq:single_queue}
\end{align}
 where $K > 0$ is a model parameter. 
The packet-drop probability $p(\cdot)$ is specified by the AQM deployed at the router. In equation~\eqref{eq:single_queue}, note the use of the notation $w(t)/\tau$ for the arrival rate instead of $x(t)$. This is because, we will now extend this model to represent the queue size evolution in the entire network routers and use it in unison with equation~\eqref{eq:gen_TCP} to form a closed-loop model for the network. 

Consider a network that comprises one edge router and $N_R$ core routers, \emph{i.e.} $N_R + 1$ routers on the whole. Note that, although we refer to these as routers without loss of generality, the model and the ensuing analysis work for any routers, switches, or network middle boxes which perform forwarding. Let $R \in \mathbb{R}^{N_R+1\times N_R+1}$ be a flow dynamics matrix where $R_{ij} \geq 0,\,\,\forall\,\, 1 \leq i,j \leq N_R+1$ represent the fraction of traffic being routed from router $i$ to router $j.$ Note that the diagonal elements satisfy $R_{ii} = 0,$ and the directionality of traffic flow implies that $R_{ij} \neq R_{ji}.$ Further, $R$ is a row substochastic matrix, with the rows with sum $< 1$ corresponding to routers from which packets may flow out of the network of routers, \emph{i.e.} routers that are connected to destination hosts. Owing to the flow dynamics matrix $R$, the number of TCP flows contributing to the total arrivals varies across the routers. Let the vector $\mathcal{T} \in \mathbb{R}^{N_R+1}$ represent the number of TCP flows being fed into the routers, and $\mathcal{C} \in \mathbb{R}^{N_R+1}$ represent the vector of service capacities of the routers. Let $q(t)\in \mathbb{R}^{N_R+1}$ represent the vector of queue sizes at the routers, as opposed to the scalar $\mathfrak{q}(t)$ in equation~\eqref{eq:single_queue}. The queue at a router $j \in [1,N_{R}+1]$, \emph{i.e.} $q_j(t)$, evolves as per
\begin{align}
\dot{q}_j(t) = \vast\{\begin{array}{l r}
             \Big(\big(1-p(t)\big)\mathcal{T}_j\frac{w(t)}{\tau} - \mathcal{C}_j\Big)\frac{1}{1+e^{-Kq_j(t)}}, & \text{if
 AQM at $j$} ,\\
             \\
              \Big(\mathcal{T}_j\frac{w(t)}{\tau} - \mathcal{C}_j\Big)\frac{1}{1+e^{-Kq_j(t)}}, & \text{otherwise}.
              \end{array}\label{eq:single_queue_approxmodel}
\end{align}
Let the edge router be labelled as $1$, then we would have $\mathcal{T}_1 = N_T,$ and the other elements of $\mathcal{T}$ can be computed using
\begin{align*}
\mathcal{T}_j = \sum_{l=1}^{N_{R}+1} R_{lj}\mathcal{T}_l, \,\,\,\forall\,\,\,\, 1 < j \leq N_R+1.
\end{align*}
The above can be expressed as the following matrix equation 
\begin{align}
\mathcal{T} = (I-R^T)^{-1}\,N_T\, \mathrm{e}_1,\label{eq:router_wise_TCPsources}
\end{align}
where $I$ is the $N_R+1\times N_R+1$ identity matrix and $\mathrm{e}_1 = [1,0,0,\cdots]^T$ is the first basis vector of the $\mathbb{R}^{N_R+1}$ space. 
Using equation~\eqref{eq:single_queue_approxmodel}, we write the system of equations that represent the evolution of queue sizes in the network of routers, with the AQM deployed at router $i$, as 
\begin{align}
\dot{q}(t) = \mathcal{F}(t)\bigg(\Big(I-\mathrm{e}_i\big(\mathcal{P}(t)\mathrm{e}_i\big)^T\Big)\mathcal{T}\,\frac{w(t)}{\tau} - \mathcal{C}\bigg),\label{eq:queue_networkmodel}
\end{align}
where $\mathcal{F}(t) \in \mathbb{R}^{N_R+1\times N_R+1}$ is a diagonal matrix with $\mathcal{F}_{jj}(t) = 1/(1+e^{-Kq_{j}(t)})$, and $\mathcal{P}(t) \in \mathbb{R}^{N_R+1\times N_R+1}$ is a diagonal matrix with $\mathcal{P}_{jj}(t) = p_j(t)$, which is the probability that a packet is dropped at router $j$. Further, $\mathrm{e}_i$ represents the $i^\text{th}$ basis vector of the $\mathbb{R}^{N_R+1}$ space. Note that the term $\mathrm{e}_i\big(\mathcal{P}(t)\mathrm{e}_i\big)^T$ produces a matrix of dimension $N_R+1 \times N_R+1$, with only one non-zero element, which is $p_i(t)$ at $i,i$. This enables us model the impact of the AQM deployed in the router $i$ on the traffic flow in the entire network. 
\subsection{Active queue management}
The threshold-based queue policy proposed in~\cite{manjunath2018stability} drops incoming packets once the queue size touches a predefined threshold. This policy is shown to ensure network stability and reduced queueing delay, while ensuring a performance that is equivalent to the Droptail queue policy in terms of throughput and packet loss. For a router that deploys the threshold-based queue policy, the packet-drop probability is given by
\begin{align}
    p(t) = \Bigg\{\begin{array}{l r}
             1 & \text{ if } \mathfrak{q}(t) \geq q_{th},\\
             \\
              0 & \text{ if } \mathfrak{q}(t) < q_{th},
              \end{array}\label{eq:packetdrop_prob_disc_threshold}
\end{align}
where $q_{th}$ is the packet-dropping threshold, which can be tuned in accordance to the incoming traffic to ensure network stability~\cite{manjunath2018stability,arxiv_version}. This packet-drop probability can be approximated using the continuous, differentiable activation function 
\begin{align}
    p(t) = \frac{1}{1+e^{-K(\mathfrak{q}(t)-q_{th})}}, \label{eq:packetdrop_prob_threshold}
\end{align}
where $K$ is the same model parameter as in equation~\eqref{eq:single_queue_approxmodel}. 
\begin{figure}[hbt!]
\centering
\scalebox{0.75}{
\begin{tikzpicture}
\begin{axis}[
yticklabel style={
/pgf/number format/fixed,
/pgf/number format/precision=2,
/pgf/number format/fixed zerofill,
/pgf/number format/fixed relative,
},
    every axis plot/.append style={thick},
    xlabel = {Queue size (pkts)},
    ylabel = {Packet-drop probability},
    xmin = 0, xmax = 910,
    ymin = 0, ymax = 1.05,
    xtick={0,250,900},
    xticklabels={$0$,$q_{th}$,$B$},
    ytick distance = 0.5,
    grid = both,
    minor tick num = 1,
    major grid style = {lightgray},
    minor grid style = {lightgray!25},
    width = 0.4\textwidth,
    height = 0.3\textwidth,
    legend cell align = {right},
    legend pos = south east
]
\addplot[blue,style= thick]  table [x = {q}, y = {p}] {figures/threshold_prob.prn};
\addplot[dashed,mark=*,red]  table [x = {q}, y = {pa}] {figures/threshold_prob.prn};
\legend{
$p(t)$ -~\eqref{eq:packetdrop_prob_disc_threshold}  , 
$p(t)$ -~\eqref{eq:packetdrop_prob_threshold}
}
\end{axis}
\end{tikzpicture}
}
\caption{Packet-drop probability at a router that deploys the threshold-based queue policy. Any packet that enters the router after the queue size has reached $q_{th}$ is dropped.} 
\label{fig:figurename}
\end{figure}
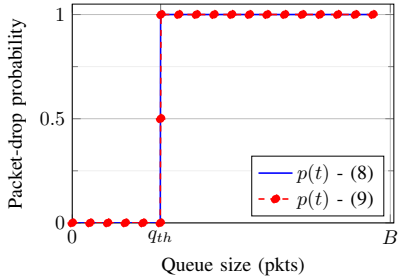

\begin{remark}
    We chose the threshold-based queue policy for two reasons: (i) it has been shown to be implementable in RED routers, and (ii) it has exactly one tunable parameter $q_{th}$ unlike other contemporary AQMs~\cite{manjunath2019compound,manjunath2018stability}. Given that our focus is on arriving at design guidelines for the deployment of AQMs and not on the optimal choice of AQM, we find it suitable to choose a simple and implementable AQM. 
\end{remark}

\begin{remark}
    Although we consider the threshold-based AQM in this work, one may replace the model for $p(t)$~\eqref{eq:packetdrop_prob_threshold} with one befitting any other queue policy to understand the impact of AQM on the packet forwarding dynamics in the network. The analytical framework used in the following sections would remain unchanged. 
\end{remark}

\subsection{Closed-loop network model}
Using the models for the average sending window of the TCP sources~\eqref{eq:gen_TCP} and the queue size evolution at the routers~\eqref{eq:queue_networkmodel}, we obtain the closed-loop model for a network of routers serving TCP flows as 
\begin{align}
\begin{split}
         \dot{w}(t) =&\, \bigg(i\big(w(t)\big)\Big(1-\mathrm{e}_i^T\big(\mathcal{P}(t-\tau)\mathrm{e}_i\big)\Big)\\
         &-d\big(w(t)\big)\mathrm{e}_i^T\big(\mathcal{P}(t-\tau)\mathrm{e}_i\big)\bigg)\frac{w(t-\tau)}{\tau},\\
    \dot{q}(t) =&\, \mathcal{F}(t)\bigg(\Big(I-\mathrm{e}_i\big(\mathcal{P}(t)\mathrm{e}_i\big)^T\Big)\mathcal{T}\,\frac{w(t)}{\tau} - \mathcal{C}\bigg),\label{eq:closed-loop_networkmodel}
\end{split}
\end{align}
where the matrix $\mathcal{P}(t)$ is now formed by the packet-drop probabilities defined by the threshold policy given by equation~\eqref{eq:packetdrop_prob_threshold}. Further, the basis vector $\mathrm{e}_i$ serves as indicator pointing to the router that deploys the threshold-based queue policy. The closed-loop system model~\eqref{eq:closed-loop_networkmodel} has the state variables $w(t)$ and $q(t)$. Recall that $q(t) \in \mathbb{R}^{N_R+1}$, hence the model comprises $N_R+2$ differential equations. Additionally, owing to the feedback delay $\tau$, the model comprises delay-differential equations which makes it infinite dimensional in nature~\cite{hale1977retarded}. 
\section{Stability Analysis}
\label{sec:Stability_Analysis}
The closed-loop model~\eqref{eq:closed-loop_networkmodel} is a non-linear, time-delayed model. For such non-linear systems, it is natural to start with analysing the system stability in a local neighbourhood of the equilibrium. Therefore, we consider a linear approximation of the above model and derive some conditions for local stability of the system about its equilibrium.
Let $(w^\ast,q^\ast)$ denote a non-trivial equilibrium of system~\eqref{eq:closed-loop_networkmodel}. Then, the equilibrium satisfies
\begin{align}
\begin{split}
i(w^\ast)(1-\mathrm{e}_i^T\mathcal{P}^\ast \mathrm{e}_i)=& \,d(w^\ast)\mathrm{e}_i^T\mathcal{P}^\ast \mathrm{e}_i,\label{eq:equilibrium}\\
\Big(I-\mathrm{e}_i(\mathcal{P}^\ast \mathrm{e}_i)^T\Big)\mathcal{T}w^\ast =&\, \mathcal{C}\tau, 
\end{split}
\end{align}
where $i(w^\ast), d(w^\ast)$ are the AIMD functions are evaluated at $w^\ast$, $\mathcal{P}^\ast$ is a diagonal matrix with $\mathcal{P}_{jj}=1/(1+e^{-K(q_j^\ast-q_{th})})$ for $1 \leq j \leq N_R+1$, which evaluates to $1$ whenever the queue at the particular router crosses the threshold $q_{th}$. Note that in the following analysis, we retain $i$ as the AQM router, and then explore the impact of the choice of $i$ on the stability of the network.
Let $\delta w(t) = w(t)-w^\ast$ and $\delta q(t) = q(t) - q^\ast$ be perturbations about the equilibrium, with $\delta q(t)$ being a vector. System~\eqref{eq:closed-loop_networkmodel} can be linearised about the equilibrium using a Taylor series expansion to obtain
\begin{equation}
\begin{aligned}
    \delta \dot{w}(t) =&\, \Big(i'(w^\ast)(1-\mathrm{e}_i^T\mathcal{P}^\ast \mathrm{e}_i) - d'(w^\ast)\mathrm{e}_i^T\mathcal{P}^\ast \mathrm{e}_i\Big)\frac{w^\ast}{\tau}\\
    &\times\delta w(t) -\Big(i(w^\ast)+d(w^\ast)\Big)\frac{w^\ast}{\tau}\mathrm{e}_i^T\mathcal{P}'^\ast \delta q(t-\tau),\\
    \delta \dot{q}(t) =&\, \frac{1}{\tau}\mathcal{F}^\ast\Big(I-\mathrm{e}_i(\mathcal{P}^\ast \mathrm{e}_i)^T\Big)\mathcal{T}\delta w(t)\\
    &- \mathrm{e}_i^T\mathcal{T}\frac{w^\ast}{\tau}\mathcal{F}^\ast \mathrm{e}_i(\mathcal{P}'^\ast \mathrm{e}_i)^T \delta q(t),\label{eq:linearised_model}
\end{aligned}
\end{equation}
where $i'(w^\ast), d'(w^\ast), \mathcal{P}'^\ast$ are first derivatives of the corresponding functions, with respect to the state variable, evaluated at equilibrium, and $\mathcal{F}^\ast$ and $\mathcal{P}^\ast$ comprise functions evaluated at equilibrium. Taking Laplace transform of system~\eqref{eq:linearised_model} yields
\begin{align*}
\lambda \delta w(\lambda) =&\, \Big(i'(w^\ast)(1-\mathrm{e}_i^T\mathcal{P}^\ast \mathrm{e}_i) - d'(w^\ast)\mathrm{e}_i^T\mathcal{P}^\ast \mathrm{e}_i\Big)\frac{w^\ast}{\tau}\delta w(\lambda) \\
&- \Big(i(w^\ast)+d(w^\ast)\Big)\frac{w^\ast}{\tau}\mathrm{e}_i^T\mathcal{P}'^\ast e^{-\lambda\tau}\delta q(\lambda),\\
\lambda \delta q(\lambda) =&\, \frac{1}{\tau}\mathcal{F}^\ast\Big(I-\mathrm{e}_i(\mathcal{P}^\ast \mathrm{e}_i)^T\Big)\mathcal{T}\delta w(\lambda) \\
&- \mathrm{e}_i^T\mathcal{T}\frac{w^\ast}{\tau}\mathcal{F}^\ast \mathrm{e}_i(\mathcal{P}'^\ast \mathrm{e}_i)^T \delta q(\lambda),
\end{align*}
which can be written as 
\begin{align*}
\begin{bmatrix}
\lambda - \mathcal{A} & -\mathcal{B}e^{-\lambda\tau}\\
-\mathcal{G} & \lambda I - \mathcal{H}
\end{bmatrix}
\begin{bmatrix}
  \delta w(\lambda)\\  
  \delta q(\lambda)
\end{bmatrix} =& \,
\mathbf{0},
\end{align*}
where 
\begin{align*}
\mathcal{A}_{1\times 1} =&\, \Big(i'(w^\ast)(1-\mathrm{e}_i^T\mathcal{P}^\ast \mathrm{e}_i) - d'(w^\ast)\mathrm{e}_i^T\mathcal{P}^\ast \mathrm{e}_i\Big)\frac{w^\ast}{\tau},\\
\mathcal{B}_{1\times N_R+1} =&\, -\Big(i(w^\ast)+d(w^\ast)\Big)\frac{w^\ast}{\tau}\mathrm{e}_i^T\mathcal{P}'^\ast,\\
\mathcal{G}_{N_R+1\times 1} =&\, \frac{1}{\tau}\mathcal{F}^\ast\Big(I-\mathrm{e}_i(\mathcal{P}^\ast \mathrm{e}_i)^T\Big)\mathcal{T},\\
\mathcal{H}_{N_R+1\times N_R+1}  =&\, - \mathrm{e}_i^T\mathcal{T}\frac{w^\ast}{\tau}\mathcal{F}^\ast \mathrm{e}_i(\mathcal{P}'^\ast \mathrm{e}_i)^T,
\end{align*}
and $\mathbf{0}$ is the zero vector of $N_R+2$ dimension. The characteristic equation of the linearised system~\eqref{eq:linearised_model} can thus be derived as 
\begin{align*}
\text{det}\begin{bmatrix}\lambda - \mathcal{A} & -\mathcal{B}e^{-\lambda\tau}\\
-\mathcal{G} & \lambda I - \mathcal{H} \end{bmatrix} =&\,0.
\end{align*}
Note that the matrix in the characteristic function is a block matrix with blocks of unequal dimensions. The determinant of which can be written as 
\begin{align*}
    (\lambda-\mathcal{A})\, \text{det}\bigg((\lambda I-\mathcal{H}) - \frac{1}{\lambda-\mathcal{A}}\,\mathcal{G}\,\mathcal{B}e^{-\lambda\tau}\bigg) =\, 0.
\end{align*}
Using the matrix determinant lemma~\cite{harville1998matrix}, the above can be simplified as 
\begin{align}
    (\lambda-\mathcal{A})\,\Big(1 - \mathcal{B}e^{-\lambda\tau}\big(\lambda I-\mathcal{H}\big)^{-1}\frac{\mathcal{G}}{\lambda-\mathcal{A}}\Big)\,\text{det}\big(\lambda I-\mathcal{H}\big) =&\, 0.
    \label{eq:matrixcharaequ}
\end{align}
Observe that the matrix $\mathcal{H}$ is a diagonal matrix. Therefore, det$(\lambda I-\mathcal{H}) = \displaystyle{\prod_{j=1}^{N_R+1}}(\lambda - \mathcal{H}_{jj})$. Now consider the second term in the characteristic equation above. Substitute $\mathcal{B}$ in this term to obtain
\begin{align*}
    1 - \mathcal{B}e^{-\lambda\tau}\big(\lambda I-\mathcal{H}\big)^{-1}\frac{\mathcal{G}}{\lambda-\mathcal{A
}} =&\, 1 + \Big(i(w^\ast)+d(w^\ast)\Big)\frac{w^\ast}{\tau}\\
&\mathrm{e}_i^T\mathcal{P}'^\ast e^{-\lambda\tau}\big(\lambda I-\mathcal{H}\big)^{-1}\frac{\mathcal{G}}{\lambda-\mathcal{A}}\\
    \end{align*}
    Upon simplification, the RHS of the above equation yields
    \begin{align*}
     =&\, 1 + \Big(i(w^\ast)+d(w^\ast)\Big)\frac{w^\ast}{\tau^2}\,\frac{\mathcal{P}'^\ast_{ii}}{(\lambda - \mathcal{H}_{ii})(\lambda-\mathcal{A})}\,e^{-\lambda\tau}\\
     &\times\,\mathcal{F}^\ast_{ii}(1-\mathcal{P}^\ast_{ii})\mathrm{e}_i^T\mathcal{T}\\
\end{align*}
Given this, the characteristic equation can be rewritten as 
\begin{align}
     &(\lambda-\mathcal{A})\,\bigg(1 + \Big(i(w^\ast)+d(w^\ast)\Big)\frac{w^\ast}{\tau}
     \,\frac{\mathcal{P}'^\ast_{ii}}{(\lambda - \mathcal{H}_{ii})(\lambda-\mathcal{A})}\,e^{-\lambda\tau}\notag\\
     &\times\,\mathcal{F}^\ast_{ii}(1-\mathcal{P}^\ast_{ii})\mathrm{e}_i^T\mathcal{T}\bigg)\,\displaystyle{\prod_{j=1}^{N_R+1}}(\lambda - \mathcal{H}_{jj})=0.\label{eq:characteristic_equation}
\end{align}
We now proceed to analyse the stability of system~\eqref{eq:linearised_model} using the characteristic equation~\eqref{eq:characteristic_equation}.
\subsection{Sufficient condition for local stability}
\label{sec:Sufficient_condition}
Any characteristic equation can be written in terms of a loop transfer function as $f(\lambda) = 1 + L(\lambda)$~\cite{aastrom2021feedback}. For the characteristic equation~\eqref{eq:characteristic_equation}, we have the loop transfer function as 
\begin{align}
    L(\lambda) =&\, \frac{\tilde{a} e^{-\lambda\tau}}{(\lambda+\tilde{b})(\lambda+\tilde{c})}, \hspace{4mm}\text{where}\label{eq:loop_transferfunction_denote}\\
    \tilde{a} =&\, \Big(i(w^\ast)+d(w^\ast)\Big)\frac{w^\ast}{\tau^2}\,\mathcal{P}'^\ast_{ii}\,\mathcal{F}^\ast_{ii}(1-\mathcal{P}^\ast_{ii})\mathrm{e}_i^T\mathcal{T},\notag\\
    \tilde{b} =&\, -\mathcal{H}_{ii},\notag \\
    \tilde{c} =&\, -\mathcal{A}.\notag
\end{align}
For the linearised system~\eqref{eq:linearised_model} to be stable, it is sufficient that the loop transfer function satisfies the Nyquist stability criteria~\cite{aastrom2021feedback}
\begin{align*}
    \angle{L(j\omega_c)} =\, \pi,\hspace{3mm}\text{and}\hspace{3mm} |L(j\omega_c)| < 1. 
\end{align*}
Using the phase condition, we obtain 
\begin{align*}
    \tan(\omega_c\tau) = \frac{\omega_c(\tilde{b}+\tilde{c})}{\omega_c^2-\tilde{b}\tilde{c}}.
\end{align*}
Using the above relationship, the magnitude condition can be rewritten as 
\begin{align}
    \frac{\tilde{a}}{\omega_c(\tilde{b}+\tilde{c})}\sin(\omega_c\tau) < 1.\label{eq:suff_cond_primary}
\end{align}
Note that $\sin(\omega_c\tau)$ attains a maximum of 1 at $\omega_c\tau = \pi/2$ in one period, it follows from~\eqref{eq:suff_cond_primary} that
\begin{align}
    \frac{\tilde{a}\tau}{\tilde{b}+\tilde{c}} < \pi/2.\label{eq:suff_cond_parameterised}
\end{align}
Substituting $\tilde{a},\tilde{b}$ and $\tilde{c}$ in equation~\eqref{eq:suff_cond_parameterised} yields
\begin{align*}
    \frac{ \Big(i(w^\ast)+d(w^\ast)\Big)\frac{w^\ast}{\tau}\,\mathcal{P}'^\ast_{ii}\,\mathcal{F}^\ast_{ii}(1-\mathcal{P}^\ast_{ii})\mathrm{e}_i^T\mathcal{T}}{ -\mathcal{H}_{ii}-\mathcal{A}} < \pi/2,
\end{align*}
which upon substituting $\mathcal{H}_{ii}$ and $\mathcal{A}$, and simplifying using equation~\eqref{eq:equilibrium} evaluates to 
    \begin{align}
    g_1(\cdot) :=\frac{d(w^\ast)\,\mathcal{F}^\ast_{ii}\mathcal{P}'^\ast_{ii}\,\mathcal{T}_i}{\mathcal{F}^\ast_{ii} \mathcal{P}'^\ast_{ii}\mathcal{T}_i-\Big(i'(w^\ast)(1-\mathcal{P}^\ast_{ii}) - d'(w^\ast)\mathcal{P}^\ast_{ii} \Big)} < \pi/2,
     \label{eq:sufficient_condition}
    \end{align}
    Inequality~\eqref{eq:sufficient_condition} serves as a sufficient condition for stability of the linearised system~\eqref{eq:linearised_model}. This implies that when condition~\eqref{eq:sufficient_condition} is satisfied, the network of routers carrying TCP flows, represented by system~\eqref{eq:closed-loop_networkmodel}, is guaranteed to be locally stable.  
    
    The function $g_1(\cdot)$ in the LHS of inequality~\eqref{eq:sufficient_condition} is a nonlinear function of the TCP increase and decrease parameters, round-trip time $\tau$, and certain parameters pertaining to the AQM router, which is denoted as $i$ here. In particular, the function depends on the equilibrium packet-drop probability at the AQM router $\mathcal{P}^\ast_{ii}$, its rate of change denoted as $\mathcal{P}'^\ast_{ii}$ and the number of flows carried by the AQM router which is given by $\mathcal{T}_i$. In order to ensure that the sufficient condition~\eqref{eq:sufficient_condition} is satisfied, one must choose an $i$ appropriately. In other words, choosing an appropriate router for AQM deployment is crucial to the stability of the network. Before exploring how the sufficient condition~\eqref{eq:sufficient_condition} can guide this choice, we present some numerical computations to understand the dependence of the sufficient condition on some of the network parameters.  
\subsubsection{Numerical computations}
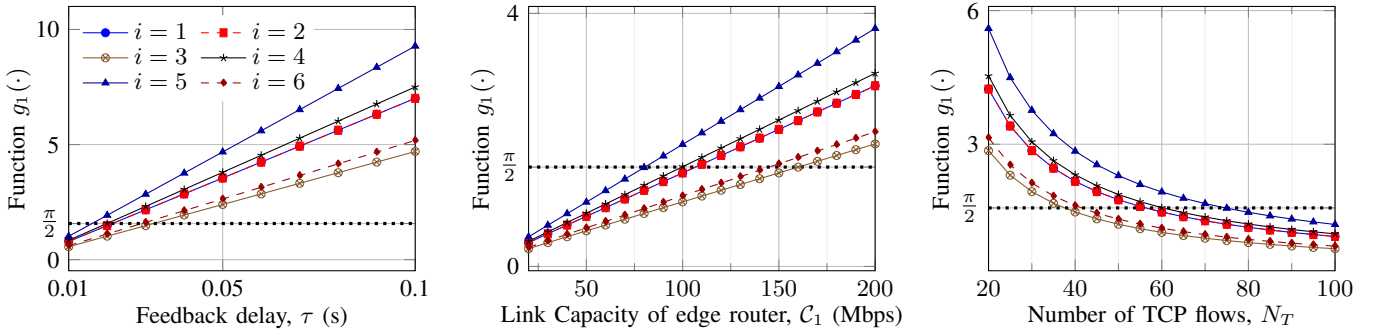
\begin{figure*}[t]
\centering
\begin{tikzpicture}

\begin{groupplot}[
  group style={
    group size=3 by 1,
    horizontal sep=1.5cm,
  },
  width=0.34\textwidth,
  height=0.28\textwidth,
  ylabel shift= -5pt,
  xlabel shift= -3pt,
  grid=both,
  minor tick num=1,
  major grid style={lightgray},
  minor grid style={lightgray!25},
  tick label style={font=\small},
  label style={font=\small},
  legend style={font=\small},
]

\nextgroupplot[
xlabel={Feedback delay, $\tau$ (s)},
ylabel={Function $g_1(\cdot)$},
ymax=11,
xmin=0.01,xmax=0.1,
ytick={0,1.57079,5,10},
yticklabels={0,\normalsize{$\frac{\pi}{2}$},5,10},
xtick={0.01,0.05,0.1},
xticklabels={0.01,0.05,0.1},
legend style={
    font=\small,
    /tikz/every even column/.append style={column sep=0.1cm},
    draw=gray!60,
    inner sep=1pt,
    row sep=-2pt,
    legend columns=2,
    legend pos=north west,
    draw=none
  },
  legend image post style={scale=1},
]

\addplot[blue, solid, mark=*, mark size=1.5pt]
table[x=tau,y=n1,col sep=space]{figures/SuffCondition_wrt_tau_NodeImpact.dat};

\addplot[red, dashed, mark=square*, mark size=1.5pt]
table[x=tau,y=n2,col sep=space]{figures/SuffCondition_wrt_tau_NodeImpact.dat};

\addplot[brown!70!black, solid, mark=otimes, mark size=1.5pt]
table[x=tau,y=n3,col sep=space]{figures/SuffCondition_wrt_tau_NodeImpact.dat};

\addplot[black, solid, mark=star, mark size=1.5pt]
table[x=tau,y=n4,col sep=space]{figures/SuffCondition_wrt_tau_NodeImpact.dat};

\addplot[blue!60!black, solid, mark=triangle*, mark size=1.5pt]
table[x=tau,y=n5,col sep=space]{figures/SuffCondition_wrt_tau_NodeImpact.dat};

\addplot[red!60!black, dashed, mark=diamond*, mark size=1.5pt]
table[x=tau,y=n6,col sep=space]{figures/SuffCondition_wrt_tau_NodeImpact.dat};

\legend{$i=1$, $i=2$, $i=3$, $i=4$, $i=5$, $i=6$}

\addplot[black, dotted, very thick]
table[x=tau,y=PI,col sep=space]{figures/SuffCondition_wrt_tau_NodeImpact.dat};

\nextgroupplot[
xlabel={Link Capacity of edge router, $\mathcal{C}_1$ (Mbps)},
ylabel={Function $g_1(\cdot)$},
ytick={0,1.57079,4},
yticklabels={0,\normalsize{$\frac{\pi}{2}$},4},
xmin=20,xmax=200
]

\addplot[blue, solid, mark=*, mark size=1.5pt]
table[x=C1,y=n1,col sep=space]{figures/SuffCondition_wrt_C1_NodeImpact.dat};

\addplot[red, dashed, mark=square*, mark size=1.5pt]
table[x=C1,y=n2,col sep=space]{figures/SuffCondition_wrt_C1_NodeImpact.dat};

\addplot[brown!70!black, solid, mark=otimes, mark size=1.5pt]
table[x=C1,y=n3,col sep=space]{figures/SuffCondition_wrt_C1_NodeImpact.dat};

\addplot[black, solid, mark=star, mark size=1.5pt]
table[x=C1,y=n4,col sep=space]{figures/SuffCondition_wrt_C1_NodeImpact.dat};

\addplot[blue!60!black, solid, mark=triangle*, mark size=1.5pt]
table[x=C1,y=n5,col sep=space]{figures/SuffCondition_wrt_C1_NodeImpact.dat};

\addplot[red!60!black, dashed, mark=diamond*, mark size=1.5pt]
table[x=C1,y=n6,col sep=space]{figures/SuffCondition_wrt_C1_NodeImpact.dat};


\addplot[black, dotted, very thick]
table[x=C1,y=PI,col sep=space]{figures/SuffCondition_wrt_C1_NodeImpact.dat};


\nextgroupplot[
xlabel={Number of TCP flows, $N_T$},
ylabel={Function $g_1(\cdot)$},
ytick={0,1.57079,3,6},
yticklabels={0,\normalsize{$\frac{\pi}{2}$},3,6},
xmin=20,xmax=100,
]

\addplot[blue, solid, mark=*, mark size=1.5pt]
table[x=NT,y=n1,col sep=space]{figures/SuffCondition_wrt_NT_NodeImpact.dat};

\addplot[red, dashed, mark=square*, mark size=1.5pt]
table[x=NT,y=n2,col sep=space]{figures/SuffCondition_wrt_NT_NodeImpact.dat};

\addplot[brown!70!black, solid, mark=otimes, mark size=1.5pt]
table[x=NT,y=n3,col sep=space]{figures/SuffCondition_wrt_NT_NodeImpact.dat};

\addplot[black, solid, mark=star, mark size=1.5pt]
table[x=NT,y=n4,col sep=space]{figures/SuffCondition_wrt_NT_NodeImpact.dat};

\addplot[blue!60!black, solid, mark=triangle*, mark size=1.5pt]
table[x=NT,y=n5,col sep=space]{figures/SuffCondition_wrt_NT_NodeImpact.dat};

\addplot[red!60!black, dashed, mark=diamond*, mark size=1.5pt]
table[x=NT,y=n6,col sep=space]{figures/SuffCondition_wrt_NT_NodeImpact.dat};


\addplot[black, dotted, very thick]
table[x=NT,y=PI,col sep=space]{figures/SuffCondition_wrt_NT_NodeImpact.dat};

\end{groupplot}
\end{tikzpicture}
\caption{Plots showing the function $g_1(\cdot)$ from the sufficient condition~\eqref{eq:sufficient_condition}. Recall that $g_1(\cdot)$ must remain below $\pi/2$ for stability. When $i=3$, \emph{i.e.} the AQM is deployed in router 3, $g_1(\cdot)$ remains below $\pi/2$ for a comparatively wider range of the parameter being varied.}\label{fig:sufficient_condition}
\end{figure*}

To begin with, we consider a network with a total of 6 routers, and the baseline values of the variable parameters are fixed at baseline values $\mathcal{C} = [100,40,40,60,60,60]^T$ Mb/s, $\tau=0.02$ s, $N_T = 60$. The flow dynamics matrix is fixed as 
\begin{align*}
    R = \begin{bmatrix}
        0 & 0.4 & 0.6 & 0 & 0 & 0\\
        0 & 0 & 0 & 0.2 & 0 & 0.6\\
        0 & 0 & 0 & 0.8 & 0.1 & 0\\
        0 & 0 & 0 & 0 & 0.7 & 0.3\\
        0 & 0 & 0 & 0 & 0 & 0.9\\
        0 & 0 & 0 & 0 & 0 & 0
    \end{bmatrix}.
\end{align*}
Note that this is representative of a network where routers have topological hierarchy, in that packets flow from the first router downstream to the remaining routers. Further, the underlying graph is not simply a tree, but has multiple acyclic paths. 
The AQM threshold is fixed at $q_{th}=100$ pkts and the model parameter $K = 100.$ Let the packet size be $1500$ bytes, with this the router capacity $\mathcal{C}_j$ can be recomputed in terms of pkts/s. For these computations, we consider Compound TCP (with default parameter values), for illustrative purposes. However, these computations can be extended to other TCP variants as well. We fix $i \in [1,6]$, to gain insight into the impact of the choice of router for AQM deployment on network stability.
With these values, we first vary the RTT in the range $\tau \in [0.01,0.1]$, and compute the function $g(\cdot)$ which appears in the LHS of condition~\eqref{eq:sufficient_condition}, for each $i$. These plots are presented in the first plot in Figure~\ref{fig:sufficient_condition}. From this plot, one can observe that as the RTT $\tau$ increases the sufficient condition gets violated, the function $g(\cdot)$. For small values $\tau$, the function $g(\cdot) < \pi/2$ thereby satisfying the inequality~\eqref{eq:sufficient_condition}. It can be seen that the curve for $i=3$ lies below the curves, and thus closest to $\pi/2$. For $i=3,$ the sufficient condition remains satisfied for larger value of $\tau$. We then vary the capacity of the edge router as $\mathcal{C}_1=[20,200]$ Mbps, the capacities of the remaining routers are scaled such that the ratios in the baseline value of $\mathcal{C}$ are maintained. From the plots obtained, we observe that a large capacity $\mathcal{C}_1$ causes a violation of the sufficient condition. Similarly, we vary the number of TCP flows as $N_T \in [20,100]$ to obtain the third plot in Fig.~\ref{fig:sufficient_condition}. This plot shows that a large number of TCP flows ensures that the sufficient condition is satisfied and the network achieves stability. All the three plots show that deploying the AQM at router $3$ ensures that the sufficient condition remains satisfied for a broader range of the variable parameter.  

\subsection{Necessary and Sufficient Condition}
The matrix function in ~\eqref{eq:matrixcharaequ} can be rewritten in a modified form that preserves the original structure of the system.
By distributing the factor $(\lambda - \mathcal{A})$ in the bracketed term and canceling the common factor in the second term, we obtain the following.
\begin{align*}
&(\lambda - \mathcal{A})
\left(
1 - \mathcal{B} e^{-\lambda \tau}
(\lambda I - \mathcal{H})^{-1}
\frac{\mathcal{G}}{\lambda - \mathcal{A}}
\right)
\\
&= (\lambda - \mathcal{A})
- \mathcal{B} e^{-\lambda \tau}
(\lambda I - \mathcal{H})^{-1}
\mathcal{G}.
\label{eq:multiplication_step}
\end{align*}
the characteristic
roots determined solely by the reduced equation.

\begin{equation}
\det(\lambda I-\mathcal{H})
\det\!\left(
\lambda-\mathcal{A}
-\mathcal{B}e^{-\lambda\tau}
(\lambda I-\mathcal{H})^{-1}\mathcal{G}
\right)
\label{eq:simplified_expression}
\end{equation}
The critical delay of the system is determined by the values $\lambda$ that satisfy
the characteristic equation. To identify the stability boundary, we substitute
$\lambda = j\omega$ into the reduced characteristic equation~\eqref{eq:simplified_expression},
given by
\begin{equation*}
\mathcal{F}(\lambda,\tau)
= \lambda - \mathcal{A}
- e^{-\lambda\tau}\,\Psi(\lambda)
= 0,
\label{eq:reduced_char}
\end{equation*}
where
\begin{equation*}
\Psi(\lambda)
= \mathcal{B}\,(\lambda I-\mathcal{H})^{-1}\mathcal{G}.
\label{eq:Psi_def}
\end{equation*}
At the stability boundary, assume the existence of purely imaginary
characteristic roots.
$\lambda = j\omega$, $\quad \omega > 0.$
Substituting $\lambda = j\omega$ into the characteristic equation yields
\[
j\omega - \mathcal{A} - e^{-j\omega\tau}\Psi(j\omega) = 0.
\]
rearranging the terms.
\[
e^{-j\omega\tau}\Psi(j\omega) = j\omega - \mathcal{A}.
\]
Let
\[
\Psi(j\omega) = \Psi_r(\omega) + j\Psi_i(\omega),
\qquad
e^{-j\omega\tau} = \cos(\omega\tau) - j\sin(\omega\tau).
\]
the real and imaginary parts of $\Psi(j\omega)$ are given by
\begin{align}
\Psi_r(\omega)
=&\,
\operatorname{Re}\{\Psi(j\omega)\}
=
-\,\mathcal{B}\,\mathcal{H}
\left(\mathcal{H}^{2}+\omega^{2}I\right)^{-1}
\mathcal{G},\label{eq:Char_RealPart}\\
\Psi_i(\omega)
=&\,
\operatorname{Im}\{\Psi(j\omega)\}
=
-\,\omega\,\mathcal{B}
\left(\mathcal{H}^{2}+\omega^{2}I\right)^{-1}
\mathcal{G}. \label{eq:Char_ImagPart}
\end{align}

Substituting these expressions and separating real and imaginary parts
leads to
\begin{align*}
-\mathcal{A}
&=
\Psi_r(\omega)\cos(\omega\tau)
+
\Psi_i(\omega)\sin(\omega\tau),
\\
\omega
&=
\Psi_i(\omega)\cos(\omega\tau)
-
\Psi_r(\omega)\sin(\omega\tau).
\end{align*}
Upon adding the squares of the above equations, we obtain 
\begin{align*}
  \mathcal{A}^2 + \omega^2 = \Psi^2_r(\omega) + \Psi^2_i(\omega) ,
\end{align*}
which after substituting equations~\eqref{eq:Char_RealPart} and~\eqref{eq:Char_ImagPart} yields the following quadratic equation in $\omega^2$
\begin{align}
   \mathcal{A}^2 + \omega^2 =&\,   \Big(\,\mathcal{B}\,\mathcal{H}
\left(\mathcal{H}^{2}+\omega^{2}I\right)^{-1}
\mathcal{G}\Big)^2\\
&+\omega^2\, \Big(\mathcal{B}
\left(\mathcal{H}^{2}+\omega^{2}I\right)^{-1}
\mathcal{G}\Big)^2\label{eq:quadratic_crossover_frequency}
\end{align}
Solving the above equation yields the frequency $\omega$ at which the characteristic roots cross over the imaginary axis. With this at hand, we now proceed to find the value of $\tau$ at which this cross over occurs. 

To eliminate trigonometric terms, the equation for the real part is multiplied by $\Psi_i(\omega)$ and the equation for the imaginary part is multiplied by $\Psi_r(\omega)$. Subtracting the resulting expressions eliminates the cosine term and yields an equation involving only $\sin(\omega\tau)$. Similarly, multiplying the real-part equation by $\Psi_r(\omega)$ and the imaginary-part equation by $\Psi_i(\omega)$, and then adding the two expressions eliminates the sine term, resulting in a complementary equation involving only $\cos(\omega\tau)$.
\begin{align*}
-\mathcal{A}\Psi_i(\omega) - \omega \Psi_r(\omega)
&=
\big(\Psi_r^2(\omega)+\Psi_i^2(\omega)\big)\sin(\omega\tau),
\\
-\mathcal{A}\Psi_r(\omega) + \omega \Psi_i(\omega)
&=
\big(\Psi_r^2(\omega)+\Psi_i^2(\omega)\big)\cos(\omega\tau).
\end{align*}
Taking the ratio of the above equations yields
\[
\tan(\omega\tau)
=
\frac{\mathcal{A}\Psi_i(\omega)+\omega\Psi_r(\omega)}
     {\mathcal{A}\Psi_r(\omega)-\omega\Psi_i(\omega)}.
\]
Denote the function on the right hand side of the above equation as $g_2(\cdot)$.
Upon substituting equations~\eqref{eq:Char_RealPart} and~\eqref{eq:Char_ImagPart} we obtain
\begin{align}
   g_2(\cdot)
=
\frac{-\omega\mathcal{B}\left(\mathcal{A}I+ \mathcal{H}\right)\left(\mathcal{H}^2+\omega^2 I\right)^{-1}\mathcal{G}}
     {\mathcal{B}\left(-\mathcal{A}\mathcal{H}+\omega^2 I\right)\left(\mathcal{H}^2+\omega^2 I\right)^{-1}\mathcal{G}}.\label{eq:Necc&Suff_function}
\end{align}
Note that all the matrices in the expression on the right hand side of the above equation are diagonal. Substituting for $\mathcal{B, A, H}$ and $\mathcal{G}$ and resolving the matrix transformation and inner product in the numerator and the denominator yields
\begin{align*}
g_2(\cdot)= \frac{-\omega\frac{w^\ast}{\tau} \Big(\big(i'(w^\ast)(1-\mathcal{P}_{ii}^{\ast}) - d'(w^\ast)\mathcal{P}_{ii}^\ast\big)-\mathcal{T}_i\mathcal{P}_{ii}^\ast \mathcal{P'}_{ii}^\ast\Big)}{\omega^2 + \left(\frac{w^\ast}{\tau}\right)^2 \big(i'(w^\ast)(1-\mathcal{P}_{ii}^{\ast}) - d'(w^\ast)\mathcal{P}_{ii}^\ast\big)\mathcal{T}_i\mathcal{P}_{ii}^\ast \mathcal{P'}_{ii}^\ast}.
\end{align*}

Hence, the delay values corresponding to purely imaginary characteristic
roots are given by
\begin{equation}
\tau_c
=
\frac{1}{\omega}
\tan^{-1}\!\left( g_2(\cdot)
\right),
\label{eq:critical_delay}
\end{equation}
where the appropriate branch of the inverse tangent function is chosen
to ensure $\tau_c > 0$. The smallest positive solution of
\eqref{eq:critical_delay} defines the \emph{critical delay} $\tau_c$, at
which stability may be lost.
Therefore, for all delays, satisfying
\begin{align}
0 \le \tau < \tau_c,\label{eq:necc_and_suff}
\end{align}
all characteristic roots remain strictly in the  left half-plane, and the equilibrium point is asymptotically stable. Hence, the above inequality provides a \emph{necessary and sufficient} condition for local stability of the closed-loop network~\eqref{eq:closed-loop_networkmodel}.
\begin{figure*}[tbh]
\centering
\begin{tikzpicture}

\begin{groupplot}[
group style={
group size= 4 by 1,
horizontal sep=1.1cm,
vertical sep=2cm
},
width=0.28\textwidth,
height=0.30\textwidth,
grid=both,
minor tick num=1,
major grid style={lightgray},
minor grid style={lightgray!25},
tick label style={font=\small},
label style={font=\small},
ylabel shift=-5pt,
legend style={draw=none, font=\small,legend columns=1}
]

\hspace{-3mm}
\nextgroupplot[
xlabel={Link capacity at router 1, $\mathcal{C}_1$ (Mbps)},
ylabel={Critical delay, $\tau_c$ (s)},
xmax=200,xmin=20,
xtick={20,80,140,200},
scaled y ticks=base 10:-2,
ytick={0, 0.05, 0.1,0.15},
yticklabels={0, 5,10,15},
ytick scale label code/.code={$\times10^{-2}$}
]

\addplot[blue, solid, mark=*, mark size=1.5pt]
table[x=C1,y=Router_1,col sep=tab]{figures/critical_delay_vs_C1.dat};

\addplot[red, dashed, mark=square*, mark size=1.5pt]
table[x=C1,y=Router_2,col sep=tab]{figures/critical_delay_vs_C1.dat};

\addplot[brown!70!black, solid, mark=otimes*, mark size=1.5pt]
table[x=C1,y=Router_3,col sep=tab]{figures/critical_delay_vs_C1.dat};

\addplot[black, solid, mark=star, mark size=1.5pt]
table[x=C1,y=Router_4,col sep=tab]{figures/critical_delay_vs_C1.dat};

\addplot[blue!60!black, solid, mark=triangle*, mark size=1.5pt]
table[x=C1,y=Router_5,col sep=tab]{figures/critical_delay_vs_C1.dat};

\addplot[red!60!black, dashed, mark=diamond*, mark size=1.5pt]
table[x=C1,y=Router_6,col sep=tab]{figures/critical_delay_vs_C1.dat};

\legend{$i=1$, $i=2$, $i=3$, $i=4$, $i=5$, $i=6$}

\nextgroupplot[
xlabel={Number of TCP flows, $N_T$},
ylabel={Critical delay, $\tau_c$ (s)},
xmax=100,xmin=20,
scaled y ticks=base 10:-2,
ytick={0,0.02,0.04},
yticklabels={0,2,4},
ytick scale label code/.code={$\times10^{-2}$}
]

\addplot[blue, solid, mark=*, mark size=1.5pt]
table[x=NT,y=Router_1,col sep=tab]{figures/critical_delay_vs_NT.dat};

\addplot[red, dashed, mark=square*, mark size=1.5pt]
table[x=NT,y=Router_2,col sep=tab]{figures/critical_delay_vs_NT.dat};

\addplot[brown!70!black, solid, mark=otimes*, mark size=1.5pt]
table[x=NT,y=Router_3,col sep=tab]{figures/critical_delay_vs_NT.dat};

\addplot[black, solid, mark=star, mark size=1.5pt]
table[x=NT,y=Router_4,col sep=tab]{figures/critical_delay_vs_NT.dat};

\addplot[blue!60!black, solid, mark=triangle*, mark size=1.5pt]
table[x=NT,y=Router_5,col sep=tab]{figures/critical_delay_vs_NT.dat};

\addplot[red!60!black, dashed, mark=diamond*, mark size=1.5pt]
table[x=NT,y=Router_6,col sep=tab]{figures/critical_delay_vs_NT.dat};

\nextgroupplot[
xlabel={Protocol parameter, $\alpha$},
ylabel={Critical delay, $\tau_c$ (s)},
ymin=0,
xmin=0.1,xmax=1,scaled y ticks=base 10:-2,
ytick={0,0.01,0.02,0.03},
yticklabels={0,1,2,3},
ytick scale label code/.code={$\times10^{-2}$}
]

\addplot[blue, solid, mark=*, mark size=1.5pt]
table[x=alpha,y=Router_1,col sep=tab]{figures/critical_delay.dat};

\addplot[red, dashed, mark=square*, mark size=1.5pt]
table[x=alpha,y=Router_2,col sep=tab]{figures/critical_delay.dat};

\addplot[brown!70!black, solid, mark=otimes*, mark size=1.5pt]
table[x=alpha,y=Router_3,col sep=tab]{figures/critical_delay.dat};

\addplot[black, solid, mark=star, mark size=1.5pt]
table[x=alpha,y=Router_4,col sep=tab]{figures/critical_delay.dat};

\addplot[blue!60!black, solid, mark=triangle*, mark size=1.5pt]
table[x=alpha,y=Router_5,col sep=tab]{figures/critical_delay.dat};

\addplot[red!60!black, dashed, mark=diamond*, mark size=1.5pt]
table[x=alpha,y=Router_6,col sep=tab]{figures/critical_delay.dat};

\nextgroupplot[
xlabel={Protocol parameter, $\beta$},
ylabel={Critical delay, $\tau_c$ (s)},
xmin=0.1,xmax=1,
scaled y ticks=base 10:-2,
ytick={0, 0.05, 0.1,0.15},
yticklabels={0, 5,10,15},
ytick scale label code/.code={$\times10^{-2}$}
]

\addplot[blue, solid, mark=*, mark size=1.5pt]
table[x=beta,y=Router_1,col sep=tab]{figures/critical_delay_vs_beta.dat};

\addplot[red, dashed, mark=square*, mark size=1.5pt]
table[x=beta,y=Router_2,col sep=tab]{figures/critical_delay_vs_beta.dat};

\addplot[brown!70!black, solid, mark=otimes*, mark size=1.5pt]
table[x=beta,y=Router_3,col sep=tab]{figures/critical_delay_vs_beta.dat};

\addplot[black, solid, mark=star, mark size=1.5pt]
table[x=beta,y=Router_4,col sep=tab]{figures/critical_delay_vs_beta.dat};

\addplot[blue!60!black, solid, mark=triangle*, mark size=1.5pt]
table[x=beta,y=Router_5,col sep=tab]{figures/critical_delay_vs_beta.dat};

\addplot[red!60!black, dashed, mark=diamond*, mark size=1.5pt]
table[x=beta,y=Router_6,col sep=tab]{figures/critical_delay_vs_beta.dat};

\end{groupplot}

\end{tikzpicture}
\caption{Variation of the critical delay $\tau_c$ with respect to link capacity $C$, the number of TCP flows $N_T$, the gain parameter $\alpha$, and parameter $\beta$ for six routers. The curves illustrate how changes in these parameters influence the stability boundary of the Compound TCP congestion control system. The legend included in the left-most plot applies to all plots.}
\label{fig:necc_suff_condition_CompoundTCP}
\end{figure*}
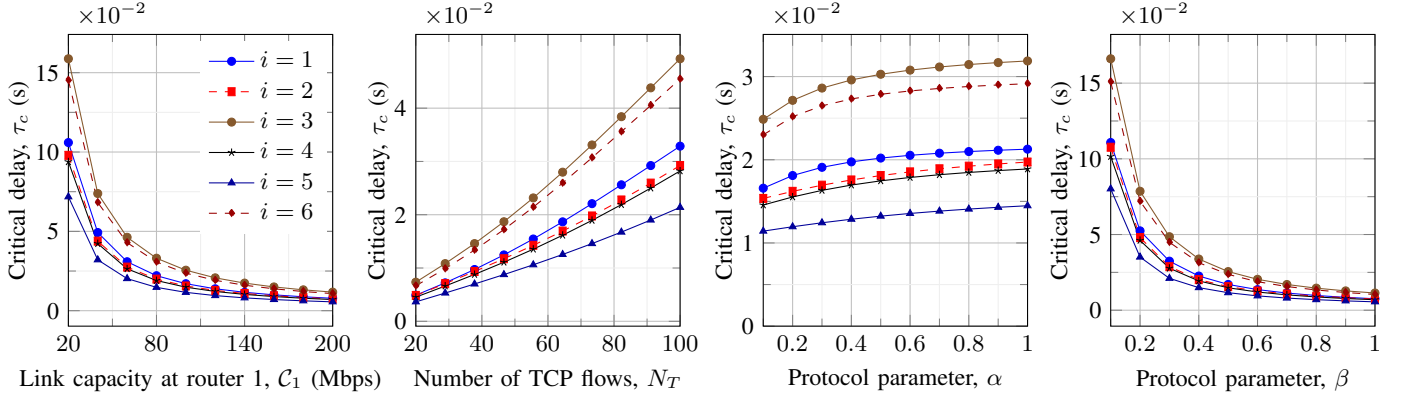
\subsubsection{Numerical computations}
We now present some computations that help to gain insight into the impact of various network and protocol parameters, as well as the choice of router for AQM deployment, on stability. The necessary and sufficient condition~\eqref{eq:necc_and_suff} enables us to define a stable region in the parameter space. In the following computations, we characterize this region in terms of the critical delay $\tau_c$ which defines the upper bound on the RTT and other network and protocol parameters. We use the same network of 6 routers and the flow dynamics matrix $R$ as in Section~\ref{sec:Sufficient_condition}. As before, we begin by understanding the impact of the service capacity of the edge router and the number of TCP flows, on system stability. The baseline values for these are fixed as $\mathcal{C} = [100,40,40,60,60,60]^T$ Mb/s and $N_T = 60$. 
Additionally, we also vary some TCP parameters to gain a deeper understanding of the stable region. This is particularly important in this case, as a violation of the necessary and sufficient condition indicates the onset of instability, unlike in the case of the sufficient condition discussed earlier, thereby necessitating an understanding of the role of protocol parameters as well. Towards this, we conduct two sets of computations, one with Compound TCP and another with DCTCP.

With Compound TCP, we first vary the capacity of the edge router $\mathcal{C}_1$ and $N_T$, with keeping one at the baseline value when the other is varied. For these computations, the Compound TCP parameters are fixed at their default values, $\alpha=0.125, k=0.75,\beta=0.5$.  The critical delay $\tau_c$ is computed using equation~\eqref{eq:critical_delay}, with the cross-over frequency $\omega$ computed using equation~\eqref{eq:quadratic_crossover_frequency}. The choice of router for AQM deployment is varied as $i \in [1,6],$ and one curve showing the relationship between $\tau_c$ and the variable parameter is obtained for $i$. These plots are shown in the top panel of Fig.~\ref{fig:necc_suff_condition_CompoundTCP}. In each plot, the region below each curve indicates the stable region. In general, we observe that as the capacity of the edge router $\mathcal{C}_1$ increases, the critical delay, \emph{i.e.} the RTT that the network can sustain without becoming unstable, reduces. This is a well-established result in networks operating TCP flows. Notably, the choice of the target router for AQM deployment $i$ impacts the stable region. Choosing $i = 3,$ expands the stable region and enable the network tolerate a larger RTT without becoming unstable. When the number of TCP flows $N_T$ increases, the network remains stable for a larger RTT. The stabilizing effect of large $N_T$ is prominent when $i=3,$ as compared to the other choices of $i.$
We then vary the Compound TCP additive increase parameter as $\alpha \in [0.1,1]$, keeping the remaining at their baseline values. The required $\tau_c$ is computed as before. Lastly, we vary the multiplicative decrease parameter $\beta \in [0.1,1].$ From these plots, we observe that varying $\alpha$ has a negligible impact on the critical delay. However, for the entire range of $\alpha$ considered, deploying the AQM at router 3 yields the largest stable region. Varying $\beta$, on the other hand, has a significant impact on network stability, with small values being favourable as they enable the network to tolerate a large RTT without the risk of instability. In this case, also, we note that $i=3$ is the best choice in terms of network stability.

\begin{figure*}[tbh]
\centering
\begin{tikzpicture}

\begin{groupplot}[
  group style={
    group size=3 by 1,
    horizontal sep=1.5cm,
  },
  width=0.34\textwidth,
  height=0.28\textwidth,
  grid=both,
  minor tick num=1,
  major grid style={lightgray},
  minor grid style={lightgray!25},
  tick label style={font=\small},
  label style={font=\small},
  legend style={font=\small},
]

\nextgroupplot[
  xlabel={Link capacity at router 1, $\mathcal{C}_1$ (Mbps)},
  ylabel={Critical delay, $\tau_c$ (s)},
  scaled y ticks=base 10:-2,
  ytick={0, 0.05, 0.10, 0.15},
  yticklabels={0, 5, 10, 15},
  ytick scale label code/.code={$\times10^{-2}$},
  xmin=20,xmax=200,
  xtick={20,80,140,200},
  legend style={
    font=\small,
    /tikz/every even column/.append style={column sep=0.1cm},
    draw=none,
    inner sep=1pt,
    row sep=-2pt,
    legend columns =2
  },
  legend image post style={scale=1},
]
\addplot[blue, solid, mark=*, mark size=1.5pt]
table[x=C1,y=Router_1,col sep=tab]{figures/critical_delay_vs_C1_DCTCP.dat};

\addplot[red, dashed, mark=square*, mark size=1.5pt]
table[x=C1,y=Router_2,col sep=tab]{figures/critical_delay_vs_C1_DCTCP.dat};

\addplot[brown!70!black, solid, mark=otimes, mark size=1.5pt]
table[x=C1,y=Router_3,col sep=tab]{figures/critical_delay_vs_C1_DCTCP.dat};

\addplot[black, solid, mark=star, mark size=1.5pt]
table[x=C1,y=Router_4,col sep=tab]{figures/critical_delay_vs_C1_DCTCP.dat};

\addplot[blue!60!black, solid, mark=triangle*, mark size=1.5pt]
table[x=C1,y=Router_5,col sep=tab]{figures/critical_delay_vs_C1_DCTCP.dat};

\addplot[red!60!black, dashed, mark=diamond*, mark size=1.5pt]
table[x=C1,y=Router_6,col sep=tab]{figures/critical_delay_vs_C1_DCTCP.dat};

\legend{$i=1$, $i=2$, $i=3$, $i=4$, $i=5$, $i=6$}

\nextgroupplot[
  xlabel={Number of flows, $N_T$},
  ylabel={Critical delay, $\tau_c$ (s)},
  xmin=20,xmax=100,
  scaled y ticks=base 10:-2,
  ytick={0, 0.02, 0.04},
  yticklabels={0, 2, 4},
  ytick scale label code/.code={$\times10^{-2}$}
]
\addplot[blue, solid, mark=*, mark size=1.5pt]
table[x=NT,y=Router_1,col sep=tab]{figures/critical_delay_vs_NT_DCTCP.dat};

\addplot[red, dashed, mark=square*, mark size=1.5pt]
table[x=NT,y=Router_2,col sep=tab]{figures/critical_delay_vs_NT_DCTCP.dat};

\addplot[brown!70!black, solid, mark=otimes, mark size=1.5pt]
table[x=NT,y=Router_3,col sep=tab]{figures/critical_delay_vs_NT_DCTCP.dat};

\addplot[black, solid, mark=star, mark size=1.5pt]
table[x=NT,y=Router_4,col sep=tab]{figures/critical_delay_vs_NT_DCTCP.dat};

\addplot[blue!60!black, solid, mark=triangle*, mark size=1.5pt]
table[x=NT,y=Router_5,col sep=tab]{figures/critical_delay_vs_NT_DCTCP.dat};

\addplot[red!60!black, dashed, mark=diamond*, mark size=1.5pt]
table[x=NT,y=Router_6,col sep=tab]{figures/critical_delay_vs_NT_DCTCP.dat};

\nextgroupplot[
  xlabel={Protocol parameter, $\bar\alpha$},
  ylabel={Critical delay, $\tau_c$ (s)},
  xmin=0.1, xmax=1,
  ymin=0, ymax=0.28,
  scaled y ticks=base 10:-2,
  ytick={0.05, 0.10, 0.15, 0.20, 0.25},
  yticklabels={5,10,15,20,25},
  ytick scale label code/.code={$\times10^{-2}$}
]
\addplot[blue, solid, mark=*, mark size=1.5pt]
table[x=alpha_bar,y=Router_1,col sep=tab]{figures/critical_delay_vs_alpha_bar_DCTCP.dat};

\addplot[red, dashed, mark=square*, mark size=1.5pt]
table[x=alpha_bar,y=Router_2,col sep=tab]{figures/critical_delay_vs_alpha_bar_DCTCP.dat};

\addplot[brown!70!black, solid, mark=otimes, mark size=1.5pt]
table[x=alpha_bar,y=Router_3,col sep=tab]{figures/critical_delay_vs_alpha_bar_DCTCP.dat};

\addplot[black, solid, mark=star, mark size=1.5pt]
table[x=alpha_bar,y=Router_4,col sep=tab]{figures/critical_delay_vs_alpha_bar_DCTCP.dat};

\addplot[blue!60!black, solid, mark=triangle*, mark size=1.5pt]
table[x=alpha_bar,y=Router_5,col sep=tab]{figures/critical_delay_vs_alpha_bar_DCTCP.dat};

\addplot[red!60!black, dashed, mark=diamond*, mark size=1.5pt]
table[x=alpha_bar,y=Router_6,col sep=tab]{figures/critical_delay_vs_alpha_bar_DCTCP.dat};

\end{groupplot}
\end{tikzpicture}
\caption{Variation of the critical delay $\tau_c$ with respect to link capacity $\mathcal{C}_1$, number of TCP flows $N_T$, and DCTCP marking parameter $\bar\alpha$ for six routers. The curves illustrate how changes in these parameters influence the stability boundary of the
Data Center TCP congestion control system. The common legend is placed in the first plot alone.}
\label{fig:necc_suff_condition_DCTCP}
\end{figure*}
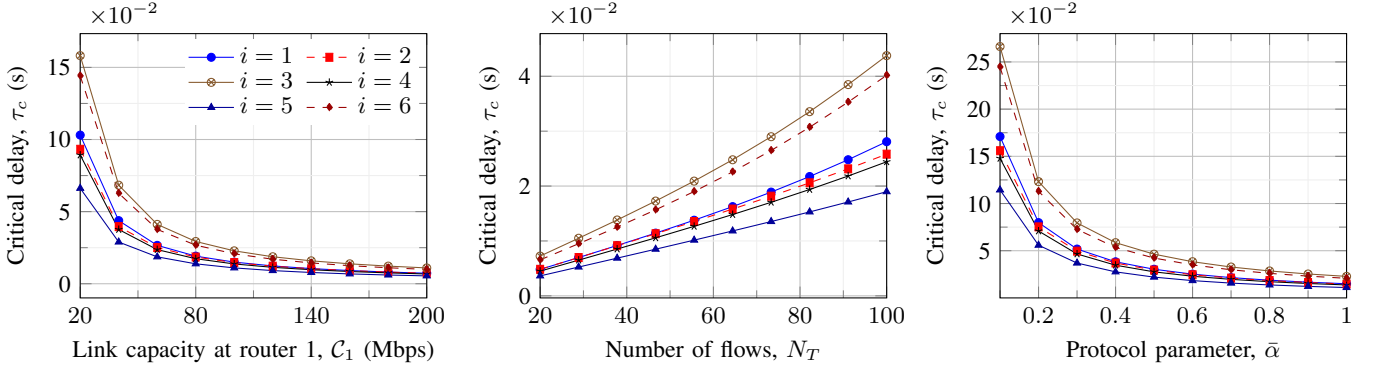

Similar computations are performed with DCTCP. These plots are presented in Fig.~\ref{fig:necc_suff_condition_DCTCP}. The first two plots, showing the impact of capacity of the edge router $\mathcal{C}_1$ and the number of TCP flows $N_T$, reveal similar insight as in the case of Compound TCP. The third plot shows the impact of varying the DCTCP protocol parameter $\bar{\alpha}$ in the range $[0.1,1]$. Observe, from the DCTCP functional form presented in Table~\ref{tab:TCP_functional_forms}, that the parameter $\bar{\alpha}$ governs the decrease function of DCTCP. The plot reveals that as $\bar{\alpha}$ increases, the network remains stable for a larger RTT. Similar to the observations made above choosing $i = 3$, has the most stabilizing effect on the network. 

\section{Guidelines for AQM deployment}
\label{sec:Centrality_Guidelines}
From the stability analysis presented above, we observed the impact of various network parameters and protocol parameters on the stability of the closed-loop network. In particular, we established that the system stability is sensitive to the choice of router for AQM deployment. We now discuss a centrality measure that enables us to quantify the importance of each router in the network.

Consider the network of routers is modelled as a directed graph, where the nodes of the graph represent the routers and each edge captures the flow of packets along the links. The fraction of flows routed from router $i$ to router $j$ is captured by the weight of the edge $(i,j)$. Consequently, the flow dynamics matrix $R$ acts as the adjacency matrix of the graph. In such directed graphs with non-negative edge weights, centrality measures serve as a metric for relative importance of the nodes~\cite{bullo2020lectures}. Many centrality measures have been proposed in the literature; some commonly used ones are in-degree centrality, eigenvector centrality, betweenness centrality. 
In our study, we employ a centrality measure known as Katz centrality to assess the relative importance of the routers, and identify the \emph{most important} router where one can deploy an AQM policy to control the congestion in the network. We find that it serves as a right measure to model various aspects that contribute to the propensity of a router to experience congestion. A discussion of this choice is presented as remarks following the definition below. 

In a directed graph with non-negative edge weights, Katz centrality of a node $j$ is given by~\cite{katz1953new,bullo2020lectures}
\begin{align}
    \mathcal{K}_j = \gamma\displaystyle\sum_{k=1}^{N_R+1}R_{kj}\mathcal{K}_k + \mu_{j},\label{eq:centrality_definition}
\end{align}
where $\gamma$ is an attenuation factor and $\mu_{j}$ represents a baseline value that captures the absolute importance of node $j$. Katz centrality is a metric that helps capture the importance of a node as the weighted sum of the $l$-length paths leading to the node, with the weights of each term being $\gamma^l.$ With $\gamma < 1,$ the largest weight is on the paths with 1 edge, and as the number of edges in the path increases, the weight decays. In general, the attenuation factor is chosen such that $\gamma < 1/\rho(R),$ where $\rho(R)$ is the absolute value of the largest eigenvalue of $R$.  Given that we intend to use Katz centrality as a measure of the routers' propensity to experience congestion, the baseline value $\mu_j$ allocated to each router must model this. To that end, we define $\mu_j = \|\mathcal{C}\|_\infty/\mathcal{C}_j \geq 1$, where $\mathcal{C}$ is the service capacity vector defined in Section~\ref{sec:QueueEvolution} and its infinity norm yields the maximum service capacity among all routers. With this definition, the baseline value assigned to each node represents the capacity scaling required at that node so that congestion would not occur. In other words, larger the required capacity scaling, higher the propensity of the router to experience congestion, and hence a large baseline value is assigned to the node. 

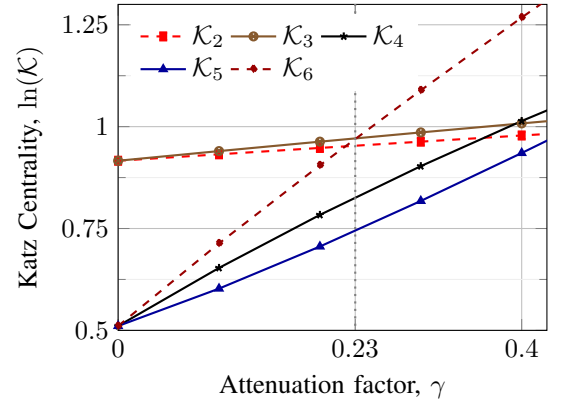
\begin{figure}[tbh]
    \centering
    \begin{tikzpicture}
    \centering
        \begin{axis}[
        yticklabel style={
        /pgf/number format/fixed,
        /pgf/number format/precision=2,
        /pgf/number format/fixed zerofill,
        /pgf/number format/fixed relative,
        },
            every axis plot/.append style={thick},
            xlabel = {Attenuation factor, $\gamma$},
            ylabel = {Katz Centrality, $\ln(\mathcal{K})$},
            xmin = 0, xmax = 0.425,
            ymin = 0.5, ymax = 1.3,
            ytick={0.5,0.75,1,1.25},
            yticklabels={$0.5$,$0.75$,$1$,$1.25$},
            xtick={0,0.235,0.4,0.8},
            xticklabels={$0$,$0.23$,$0.4$,$0.8$},
            grid = both,
            minor tick num = 1,
            major grid style = {lightgray},
            minor grid style = {lightgray!25},
            width = 0.4\textwidth,
            height = 0.325\textwidth,
            legend cell align = {right},
            legend pos = north west,
            legend columns=3,
            legend style={draw=none}
        ]
        \addplot[red, dashed, mark=square*, mark size=1.5pt]  table [x = {g}, y expr={ln(\thisrow{n2})}] {figures/CentralityValues.dat};
        \addplot[brown!70!black, solid, mark=otimes, mark size=1.5pt]  table [x = {g}, y expr={ln(\thisrow{n3})}] {figures/CentralityValues.dat};
        \addplot[black, solid, mark=star, mark size=1.5pt]  table [x = {g},y expr={ln(\thisrow{n4})}] {figures/CentralityValues.dat};
        \addplot[blue!60!black, solid, mark=triangle*, mark size=1.5pt]  table [x = {g}, y expr={ln(\thisrow{n5})}] {figures/CentralityValues.dat};
        \addplot[red!60!black, dashed, mark=diamond*, mark size=1.5pt]  table [x = {g}, y expr={ln(\thisrow{n6})}] {figures/CentralityValues.dat};
        \draw[dotted,gray,thick] (axis cs:0.235,0) -- (axis cs:0.235,10);
        \legend{
        $\mathcal{K}_2$,
        $\mathcal{K}_3$, 
        $\mathcal{K}_4$,
        $\mathcal{K}_5$,
        $\mathcal{K}_6$
        }
        \end{axis}
    \end{tikzpicture}
    \caption{Plot showing Katz centrality measures, in log scale, of routers 2 through 6 computed using various values of the attenuation factor $\gamma$. Observe that for $\gamma < 0.23,$ router 3 has the highest centrality, and for $\gamma > 0.23$ the centrality of router 6 increases beyond that of router 3.}
    \label{fig:centrality_values}
\end{figure}

To better understand Katz centrality and its dependence on the attenuation factor, we conducted computations using flow dynamics matrix $R$ and the service capacity vector $\mathcal{C}$ used in the computations presented in Section~\ref{sec:Stability_Analysis}. For this choice of $\mathcal{C}$, we have $\|\mathcal{C}\|_\infty=100$. Therefore, the baseline values $\mu_j$ for all $j \in \{1,\cdots,6\}$ is chosen using $\mu_j = 100/\mathcal{C}_j$. The centrality measures computed for various values of the attenuation $\gamma$, for all $j \in \{2,\cdots,6\}$ are plotted in Fig.~\ref{fig:centrality_values}. The first router has the maximum capacity, and hence has a baseline value of $\mu_1 = 1.$ Further, it has no incoming edges from any other routers. Therefore, it has a constant Katz centrality of $1,$ regardless of the value of $\gamma,$ and is not shown in the plot. As the value of $\gamma$ increases, the centrality values of all routers increase. Further, we observe that when $\gamma < 0.23$, router 3 has the largest centrality and when $\gamma$ increases further, router 6 has the largest centrality. Note that router 3 has only a 1-length path, while router 6 has multiple 2- and 3-length paths. Recall that an $l$-length path is given a weight of $\gamma^l.$ As $\gamma$ increases, the weight given to the 2- and 3-length paths to router 6 also increases, thereby making it the most central router. 

Now, to understand the impact of Katz centrality on network stability, we revisit the sufficient condition and the necessary and sufficient condition for local stability derived in Section~\ref{sec:Stability_Analysis}. Simplifying the function $g_1(\cdot)$ in the sufficient condition~\eqref{eq:sufficient_condition} using the equilibrium condition~\eqref{eq:equilibrium} gives
\begin{align*}
g_1(\cdot)=\frac{\Big(i(w^\ast)+d(w^\ast)\Big)\mathcal{C}_i \tau \mathcal{F}^\ast_{ii}\mathcal{P}'^\ast_{ii}}{\Delta},
\end{align*}
where $\Delta = w^\ast\Big(\mathcal{F}^\ast_{ii} \mathcal{P}'^\ast_{ii}\mathcal{T}_i-i'(w^\ast)(1-\mathcal{P}^\ast_{ii}) + d'(w^\ast)\mathcal{P}^\ast_{ii}\Big).$
Substituting for $\mathcal{C}_i$ using equation~\eqref{eq:centrality_definition}, we obtain
\begin{small}
\begin{align*}
 g_1(\cdot)=   \frac{\Big(i(w^\ast)+d(w^\ast)\Big)\|\mathcal{C}\|_\infty \tau \mathcal{F}^\ast_{ii}\mathcal{P}'^\ast_{ii}}{\Big(\mathcal{K}_i - \gamma\displaystyle\sum_{k=1}^{N_R+1}R_{ki}\mathcal{K}_k\Big)\Delta}
\end{align*}
\end{small}

Next, consider the function $g_2(\cdot)$, defined in equation~\eqref{eq:Necc&Suff_function}. Simplifying the function and using it in the necessary and sufficient condition~\eqref{eq:critical_delay} yields
\begin{align*}
   \tau_c 
   =&\, \frac{1}{\omega}\tan^{-1}\bigg(\frac{\mathcal{H}_{ii}}{\omega}\bigg) -  \frac{1}{\omega} \tan^{-1}\bigg(\frac{\mathcal{A}}{\omega}\bigg).
\end{align*}
Denote the two components of $\tau_c$ as 
\begin{align*}
    \tau_{c1} :=  \frac{1}{\omega}\tan^{-1}\bigg(\frac{\mathcal{H}_{ii}}{\omega}\bigg),  && \tau_{c2} := \frac{1}{\omega} \tan^{-1}\bigg(\frac{\mathcal{A}}{\omega}\bigg).
\end{align*}
We first consider $\tau_{c1}$.
Substituting $H_{ii}$, and simplifying using the equilibrium condition~\eqref{eq:equilibrium} we obtain 
\begin{align*}
   \tan(\omega\,  \tau_{c1}) = \frac{-\, \mathcal{C}_i \,\mathcal{F}^\ast_{ii}\, \mathcal{P'}^\ast_{ii}} {\omega\, (1-\mathcal{P}^\ast_{ii})}.
\end{align*}
Now, we substitute for $\mathcal{C}_i$ using equation~\eqref{eq:centrality_definition}, which gives
\begin{align}
    \tau_{c1} =&\, \frac{-1}{\omega} \tan^{-1}\left(\frac{ \|\mathcal{C}\|_\infty \,\mathcal{F}^\ast_{ii}\, \mathcal{P'}^\ast_{ii}} {\omega\, (1-\mathcal{P}^\ast_{ii})\,\big( \mathcal{K}_i - \gamma \displaystyle{\sum_{k=1}^{N_R+1}}R_{ki}\mathcal{K}_k \big)\,}\right)
\end{align}
 Observe that $\tau_{c1}$ increases as the Katz centrality of router $i$, $\mathcal{K}_i$, increases. The other component of the critical delay,  $\tau_{c2}$, does not depend explicitly on the choice of $i$, as the term $\mathcal{A}$ can be shown to be independent of $i$ using the equilibrium condition~\eqref{eq:equilibrium}. 
 Therefore, based on the above observation, the choice of $i$ can be given by $i := \argmax_i \mathcal{K}_i$. This implies that one may choose the router with the highest Katz centrality for AQM deployment. This matches the observations from the numerical computations. From Fig.~\ref{fig:centrality_values}, we observe that router 3 has the largest centrality when $\gamma < 0.23,$ and router 6 is more central when $\gamma > 0.23.$ Indeed from the plots presented in Figs.~\ref{fig:sufficient_condition},~\ref{fig:necc_suff_condition_CompoundTCP},~\ref{fig:necc_suff_condition_DCTCP} we observe that the network remains stable for comparatively larger RTT when the AQM is deployed at router 3, with router 6 being the next best choice. Although, either router 3 or router 6 can be regarded as the router with largest Katz centrality, based on the choice of $\gamma$, we see that deploying the AQM at router 3 has a greater stabilizing effect. It is important to understand the correlation between the choice of $\gamma$ and the stabilizing effect of deploying the AQM. For $\gamma > 0.23$, according to the proposed Katz centrality metric, an AQM must be deployed in router 6. This causes congestion to percolate deeper into the network, making it more vulnerable to instability due to large RTT compared to the case where router 3 has the AQM. This indicates that while deploying the AQM at the router with largest Katz centrality has a stabilizing effect on the network, the centrality itself must ideally be computed using a small attenuation factor to reap the greatest benefit of the proposed design.

\begin{remark}
For the network used in the computations above, edge-based AQM deployment proposed in~\cite{zhu2006edge,briscoe2023low} implies that the AQM would be deployed in router 1, \emph{i.e.} $i=1$. From the plots presented in Figs.~\ref{fig:sufficient_condition},~\ref{fig:necc_suff_condition_CompoundTCP},~\ref{fig:necc_suff_condition_DCTCP}, it can be observed that deploying AQM at router 1 has the least stabilizing effect on the network, among all possible choices. Clearly, the proposed design of choosing the router with the largest Katz centrality, outperform edge-based AQM deployment. 
\end{remark}
\begin{remark}
    Recall that rows of $R$ that have a sum $< 1$ correspond to routers connected to destination hosts. For the matrix $R$ used in the computations, row 6 has the smallest row sum, and hence router 6 would be connected to the largest number of destination hosts. A simplistic viewpoint would dictate that deploying an AQM at this router would be optimal. However, the above centrality-based design dictates otherwise, and is supported by the stability analysis presented above. 
\end{remark}
\section{Packet-level simulations}
\label{sec:Simulations}

In this section, we present packet-level simulations conducted ns-3 (version 3.38). Through these simulations, we conducted a performance evaluation of an illustrative multi-bottleneck topology with acyclic paths, which mimics the conditions of the network considered in the computations presented in Section~\ref{sec:Stability_Analysis}. While the analysis and computations presented above establish that using Katz centrality a metric for AQM deployment proves beneficial in terms of network stability, the following simulation-based study is aimed at evaluating the latency performance of the network.  

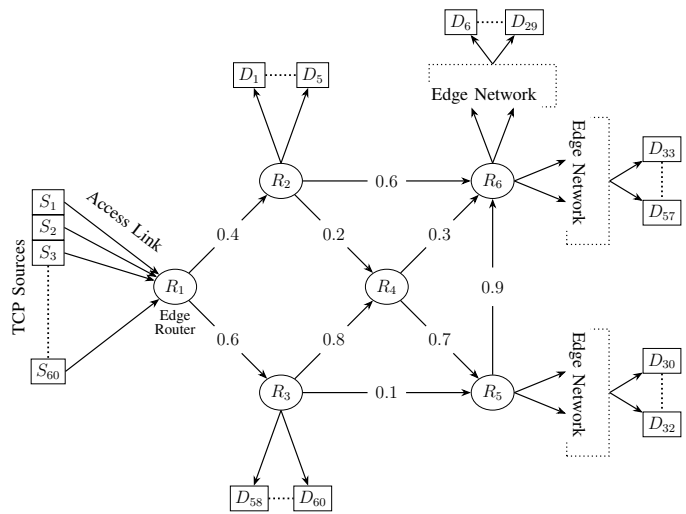
\begin{figure}[tbh]
\begin{center}
\scalebox{0.56}{
\begin{tikzpicture}[->]
    \node[ellipse,draw,font=\large] (R1) at (10,10) {$R_1$};
    \node (AR1) at (10,9.3) {Edge};
    \node (AR2) at (10,9) {Router};
    \node[ellipse,draw,font=\large] (R2) at (12.5,12.5) {$R_2$};
    \node[ellipse,draw,font=\large] (R3) at (12.5,7.5) {$R_3$};
    \node[ellipse,draw,font=\large] (R4) at (15,10) {$R_4$};
    \node[ellipse,draw,font=\large] (R6) at (17.5,12.5) {$R_6$};
    \node[ellipse,draw,font=\large] (R5) at (17.5,7.5) {$R_5$};
    \node[rectangle,draw,minimum width=7.5mm,font=\large] (S1) at (7,12) {$S_1 $};
    \node[rectangle,draw,minimum width=7.5mm,font=\large] (S2) at (7,11.4) {$S_2 $};
    \node[rectangle,draw,minimum width=7.5mm,font=\large] (S3) at (7,10.8) {$S_3 $};
    \node[rectangle,draw,minimum width=7.5mm,font=\large] (S60) at (7,8) {$S_{60}$};
    \node[rectangle,draw,minimum width=4mm,font=\large] (D1) at (11.75,15) {$D_{1}$};
    \node[rectangle,draw,minimum width=7.5mm,font=\large] (D5) at (13.25,15) {$D_{5}$};
    \node[rectangle,draw,thick,minimum width=3cm,minimum height=1cm,dotted] (AN2) at (17.5,14.75){};
    \node[rectangle,draw,minimum width=7.5mm,font=\large,font=\large] (D6) at (16.75,16.25) {$D_{6}$};
    \node[rectangle,draw,minimum width=7.5mm,font=\large,font=\large] (D29) at (18.25,16.25) {$D_{29}$};
    \node[rectangle,draw,thick,minimum height=3cm,minimum width=1cm,dotted] (AN3) at (19.75,12.5){};
    \node[rectangle,draw,thick,minimum height=3cm,minimum width=1cm,dotted] (AN4) at (19.75,7.5){};
    \node[rectangle,draw,minimum width=7.5mm,font=\large] (D33) at (21.5,13.25) {$D_{33}$};
    \node[rectangle,draw,minimum width=7.5mm,font=\large] (D57) at (21.5,11.75) {$D_{57}$};
    \node[rectangle,draw,minimum width=7.5mm,font=\large] (D30) at (21.5,8.25) {$D_{30}$};
    \node[rectangle,draw,minimum width=7.5mm,font=\large] (D32) at (21.5,6.75) {$D_{32}$};
    \node[rectangle,draw,minimum width=7.5mm,font=\large] (D58) at (11.75,5) {$D_{58}$};
    \node[rectangle,draw,minimum width=7.5mm,font=\large] (D60) at (13.25,5) {$D_{60}$};
    \node [rectangle, draw=white,fill=white, minimum width=2cm,rotate=90,font=\large] (TS) at (6.3,10) {TCP Sources};
    \node[rectangle,draw=white,fill=white,font=\large] at (17.3,14.5) {Edge Network};
    \node[rectangle,draw=white,fill=white,rotate=270,font=\large] at (19.5,12.7) {Edge Network};
    \node[rectangle,draw=white,fill=white,rotate=270,font=\large] at (19.5,7.7) {Edge Network};
   \begin{scope}[>={Stealth[black]},
          every node/.style={fill=white, circle},
          every edge/.style={draw=black ,thick}] 
    \path  [->]  (R1) edge node {\large{$0.4$}} (R2);
    \path [->]  (R1) edge  node {\large{$0.6$}} (R3);
    \path [->]  (R2) edge  node {\large{$0.2$}} (R4);
    \path [->]  (R2) edge  node {\large{$0.6$}} (R6);
    \path [->]  (R3) edge  node {\large{$0.8$}} (R4);
    \path [->]  (R3) edge  node {\large{$0.1$}} (R5);
    \path [->]  (R4) edge  node  {\large{$0.7$}} (R5);
    \path [->]  (R4) edge  node  {\large{$0.3$}} (R6);
    \path [->]  (R5) edge  node  {\large{$0.9$}} (R6);
    \path  [->]  (S1.east) edge node[midway,above,sloped,yshift=-0.7cm,fill=none] {\large{Access Link}} (R1);
    \path [->] (S2.east) edge (R1);
    \path [->] (S3.east) edge (R1);
    \path [->] (S60.east) edge (R1);
    \path [->] (R2.north) edge (D1);
    \path [->] (R2.north) edge (D5);
    \path [->] (R6.north) edge ([xshift=-0.5cm]AN2.south);
    \path [->] (R6.north) edge ([xshift=0.5cm]AN2.south);
    \path [->] (AN2.north) edge (D6);
    \path [->] (AN2.north) edge (D29);
    \path [->] (R6.east) edge ([yshift=-0.5cm]AN3.west);
    \path [->] (R6.east) edge ([yshift=0.5cm]AN3.west);
    \path [->] (R5.east) edge ([yshift=-0.5cm]AN4.west);
    \path [->] (R5.east) edge ([yshift=0.5cm]AN4.west);
    \path [->] (AN3.east) edge (D33);
    \path [->] (AN3.east) edge (D57);
    \path [->] (AN4.east) edge (D30);
    \path [->] (AN4.east) edge (D32);
    \path [->] (R3.south) edge (D58);
    \path [->] (R3.south) edge (D60);
    \end{scope}
    \begin{scope}[>={Stealth[black]},
          every edge/.style={draw=black ,dotted,very thick}] 
     \path [-] (S3) edge (S60); 
     \path [-] (D1) edge (D5); 
     \path [-] (D6) edge (D29); 
     \path [-] (D30) edge (D32); 
     \path [-] (D33) edge (D57);
     \path [-] (D58) edge (D60);
    \end{scope}
  \end{tikzpicture}
  }
\end{center}
\caption{Schematic representation of the network topology used for packet-level simulations. TCP sources $S_1, S_2, \cdots S_{60}$ are connected to destination hosts $D_1, D_2, \cdots D_{60}$, respectively. All the sources feed into the edge router $R_1$, the destination hosts are spread across the network as shown. The weight of each edge depicts the fraction of traffic routed along the edge.}
\label{fig:topology}
\end{figure}
A schematic diagram of topology considered is presented in Fig.~\ref{fig:topology}. This topology has the same structure as the network considered for the computations in Section~\ref{sec:Stability_Analysis}. The network has a total of 60 TCP sources-destination pairs and 6 routers, with all the sources connected to the edge router, enumerated as 1. The capacity of the outgoing links of the routers is fixed as $\mathcal{C}=[100,40,40,60,60]$ Mbps, these links are configured to have negligible propagation delay. All sources and destinations are connected to the network with dedicated access links of 2 Mbps capacity. The buffers at the routers are sized according to the bandwidth-delay product thumb rule, using an RTT of 250 ms which is generally used for buffer sizing~\cite{raina2005buffer}. To reflect the substochasticity of the flow dynamics matrix, corresponding destinations (outflows) have been added to Routers 2, 3, 5 and 6 for adjustment. Packets are routed using the \texttt{Ipv4StaticRouting} helper, to ensure that the traffic flow in the network mimics the chosen matrix $R$. The threshold parameter $q_{th}$ of the AQM is fixed at 100 pkts. 

In this setup, we choose two variants of TCP; TCP NewReno and DCTCP. The packet size is fixed as 1500 Bytes. As the overarching goal is to explore the impact of increased RTT, we vary the propagation delay in the range $[10,300]$ ms to force an increase in RTT. For each value of propagation delay in the given range, say $T$, the delay between each source destination pair is chosen randomly as $X \sim \mathcal{U}[T-5,T+5],$ and this is fixed by setting the one-way delay of each access link as $X/4.$ Such a randomisation ensures that the TCP flows are not synchronised due to the same RTT. Overall, the network would have a mean propagation delay of $T$, while the propagation delay across TCP flows varies. 
Each simulation is run for a total of 250 seconds, ensuring that the network reaches a steady-state. All TCP flows are started at a randomly chosen time instant within the first 5 seconds of the simulation, and they remain until the simulation ends. For comparison, we simulate two scenarios: (i) without any AQM, \emph{i.e} all DropTail routers, (ii) with the threshold-based AQM deployed at router 3, and the remaining routers use DropTail. 

\begin{figure*}[tbh!]
    \centering
    \begin{tikzpicture}
        \def\ycol{avgLatency}
        \def\yerrcol{stdLatency}
        \def\ylabel{Latency (ms)}
        \def\ymin{5}        
        \begin{axis}[
            xlabel={Propagation delay (ms)},
            ylabel={\ylabel},
            xmin=10, xmax=300,
            ymin=-20,ymax=650,
            xtick={10, 50, 100, 150, 200, 250, 300},
            legend pos=south east,
            legend style={font=\scriptsize,draw=none},
            grid=both,
            major grid style={dashed,line width=0.5pt,draw=gray!50},
            minor grid style={dotted,line width=0.25pt,draw=gray!25},
            yticklabel style={/pgf/number format/fixed},
            x tick label style={rotate=45,anchor=east},
            width=2.2in, height=2in,
        ]
        \addplot[
            color=red, mark=*, mark options={fill=red},
            error bars/.cd, y dir=both, y explicit, error bar style={line width=0.8pt},
        ] table[x=RTT, y=\ycol, y error=\yerrcol, col sep=comma,each nth point=3]{figures/metrics_AQMn3_DCTCP.csv};
        \addlegendentry{AQM at router 3}
        \addplot[
            color=blue, mark=square*, mark options={fill=blue},
            error bars/.cd, y dir=both, y explicit, error bar style={line width=0.8pt},
        ] table[x=RTT, y=\ycol, y error=\yerrcol, col sep=comma,each nth point=3] {figures/metrics_noAQM_DCTCP.csv};
        \addlegendentry{DropTail}
        \end{axis}
    \end{tikzpicture}
    \quad
    \begin{tikzpicture}
        \def\ycol{avgPacketLoss}
        \def\yerrcol{stdPktLoss}
        \def\ylabel{Packet Loss (Bytes)}
        \def\ymin{-20000}
        
        \begin{axis}[
            xlabel={Propagation delay (ms)},
            ylabel={\ylabel},
            xmin=10, xmax=300,
            ymin=\ymin,
            xtick={10, 50, 100, 150, 200, 250, 300},
            legend pos=north east,
            grid=both,
            major grid style={dashed,line width=0.5pt,draw=gray!50},
            minor grid style={dotted,line width=0.25pt,draw=gray!25},
            yticklabel style={/pgf/number format/fixed},
            x tick label style={rotate=45,anchor=east},
            width=2.2in, height=2in,
        ]
        \addplot[
            color=red, mark=*, mark options={fill=red},
            error bars/.cd, y dir=both, y explicit, error bar style={line width=0.8pt},
        ] table[x=RTT, y=\ycol, y error=\yerrcol, col sep=comma,each nth point=3]{figures/metrics_AQMn3_DCTCP.csv};
        \addplot[
            color=blue, mark=square*, mark options={fill=blue},
            error bars/.cd, y dir=both, y explicit, error bar style={line width=0.8pt},
        ] table[x=RTT, y=\ycol, y error=\yerrcol, col sep=comma,each nth point=3] {figures/metrics_noAQM_DCTCP.csv};
        \end{axis}
    \end{tikzpicture}
   \quad
    \begin{tikzpicture}
        \def\ycol{avgThroughput}
        \def\yerrcol{stdThroughput}
        \def\ylabel{Throughput (Mbps)}
        \def\ymin{-0.5}
        
        \begin{axis}[
            xlabel={Propagation delay (ms)},
            ylabel={\ylabel},
            xmin=10, xmax=300,
            ymin=\ymin, ymax=12,
            xtick={10, 50, 100, 150, 200, 250, 300},
            ytick={0,4,8,12},
            legend pos=south west,
            grid=both,
            major grid style={dashed,line width=0.5pt,draw=gray!50},
            minor grid style={dotted,line width=0.25pt,draw=gray!25},
            yticklabel style={/pgf/number format/fixed},
            x tick label style={rotate=45,anchor=east},
            width=2.2in, height=2in,
        ]
        \addplot[
            color=red, mark=*, mark options={fill=red},
            error bars/.cd, y dir=both, y explicit, error bar style={line width=0.8pt},
        ] table[x=RTT, y=\ycol, y error=\yerrcol, col sep=comma,each nth point=3] {figures/metrics_AQMn3_DCTCP.csv};
        \addplot[
            color=blue, mark=square*, mark options={fill=blue},
            error bars/.cd, y dir=both, y explicit, error bar style={line width=0.8pt},
        ] table[x=RTT, y=\ycol, y error=\yerrcol, col sep=comma,each nth point=3] {figures/metrics_noAQM_DCTCP.csv};
        \end{axis}
    \end{tikzpicture} 
    \caption{Performance evaluation metrics across varying network propagation delays for DCTCP variant. The common legend appears in the left plot alone.}
    \label{fig:perf_eval_results_DCTCP}
\end{figure*}
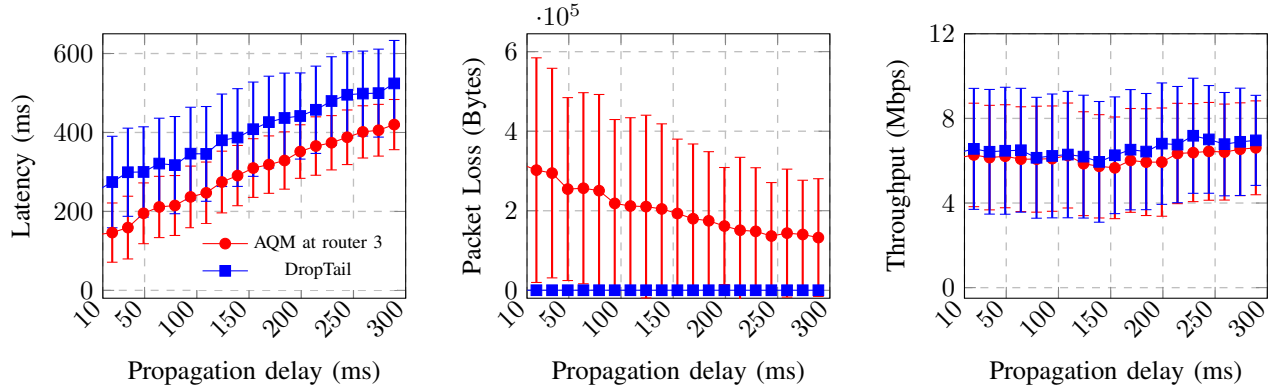

For each simulation, we record three key metrics: (i) throughput, (ii) packet loss, and (iii) latency. Throughput is recorded for each TCP flow, using the \texttt{FlowMonitor} helper. Packet loss is recorded at each router using the \texttt{TxQueue/Drop} trace source, and the total packet loss is calculated. For latency, we trace the occupancy at each router queue by querying the size of the \texttt{QueueDisc} and then scale it by the capacity of the outgoing link to calculate the queueing delay. Adding the queueing delay at each router enroute to the propagation delay yields the latency. Once these metrics are traced for each simulation, we calculated the mean and standard deviation of the last 50 seconds of the traces which reflect the steady state performance of the network. These plots are shown in Fig.~\ref{fig:perf_eval_results_NewReno} for NewReno and Fig.~\ref{fig:perf_eval_results_DCTCP} for DCTCP.

In case of NewReno, we observe that deploying the AQM at router 3 leads to a  reduction in latency. The reduced deviation in latency implies reduced jitter and indicates improved user experience. Deploying the threshold-based AQM causes router 3 to deliberately drop packets tp keep the queue size small. This is reflected in the packet loss, where we see the DropTail queues drop no packets at all, while the network with the AQM drops a large number of packets. However, the packet loss reduces as the propagation delay increases. Given that the network drops a substantial amount of packets when the AQM is implemented, one would expect the throughput to drop. However, this does not happen. The AQM drops packets to keep the queue size low, but the lost packets do not impact the throughput as the latency is substantially reduced. The observations on latency and packet loss are seen to hold in case of DCTCP as well. However, with DCTCP, the throughput is reduced substantially when the AQM is implemented. A more detailed investigation with a traffic composition included other TCP variants, along with UDP flows and short bursty traffic, and a wider variety of performance metrics is in order.


\begin{figure*}[tbh!]
    \centering
    \begin{tikzpicture}
        \def\ycol{avgLatency}
        \def\yerrcol{stdLatency}
        \def\ylabel{Latency (ms)}
        \def\ymin{5}
        
        \begin{axis}[
            xlabel={Propagation delay (ms)},
            ylabel={\ylabel},
            xmin=10, xmax=300,
            ymin=-20,ymax=650,
            xtick={10, 50, 100, 150, 200, 250, 300},
            legend pos=south east,
            legend style={font=\scriptsize,draw=none},
            grid=both,
            major grid style={dashed,line width=0.5pt,draw=gray!50},
            minor grid style={dotted,line width=0.25pt,draw=gray!25},
            yticklabel style={/pgf/number format/fixed},
            x tick label style={rotate=45,anchor=east},
            width=2.2in, height=2in,
        ]
        \addplot[
            color=red, mark=*, mark options={fill=red},
            error bars/.cd, y dir=both, y explicit, error bar style={line width=0.8pt},
        ] table[x=RTT, y=\ycol, y error=\yerrcol, col sep=comma,each nth point=3] {figures/metrics_AQMn3_Reno.csv};
        \addlegendentry{AQM at router 3}
        \addplot[
            color=blue, mark=square*, mark options={fill=blue},
            error bars/.cd, y dir=both, y explicit, error bar style={line width=0.8pt},
        ] table[x=RTT, y=\ycol, y error=\yerrcol, col sep=comma,each nth point=3] {figures/metrics_noAQM_Reno.csv};
        \addlegendentry{DropTail}
        \end{axis}
    \end{tikzpicture}
    \quad
    \begin{tikzpicture}
        \def\ycol{avgPacketLoss}
        \def\yerrcol{stdPktLoss}
        \def\ylabel{Packet Loss (Bytes)}
        \def\ymin{-2000}  
        \begin{axis}[
            xlabel={Propagation delay (ms)},
            ylabel={\ylabel},
            xmin=10, xmax=300,
            ymin=\ymin,
            xtick={10, 50, 100, 150, 200, 250, 300},
            legend pos=north east,
            grid=both,
            major grid style={dashed,line width=0.5pt,draw=gray!50},
            minor grid style={dotted,line width=0.25pt,draw=gray!25},
            yticklabel style={/pgf/number format/fixed},
            x tick label style={rotate=45,anchor=east},
            width=2.2in, height=2in,
        ] 
        \addplot[
            color=red, mark=*, mark options={fill=red},
            error bars/.cd, y dir=both, y explicit, error bar style={line width=0.8pt},
        ] table[x=RTT, y=\ycol, y error=\yerrcol, col sep=comma,each nth point=3] {figures/metrics_AQMn3_Reno.csv};
        \addplot[
            color=blue, mark=square*, mark options={fill=blue},
            error bars/.cd, y dir=both, y explicit, error bar style={line width=0.8pt},
        ] table[x=RTT, y=\ycol, y error=\yerrcol, col sep=comma,each nth point=3] {figures/metrics_noAQM_Reno.csv};
        \end{axis}
    \end{tikzpicture}
    \quad
    \begin{tikzpicture}
        \def\ycol{avgThroughput}
        \def\yerrcol{stdThroughput}
        \def\ylabel{Throughput (Mbps)}
        \def\ymin{-0.5}
        \begin{axis}[
            xlabel={Propagation delay (ms)},
            ylabel={\ylabel},
            xmin=10, xmax=300,
            ymin=\ymin, ymax=12,
            ytick={0,4,8,12},
            xtick={10, 50, 100, 150, 200, 250, 300},
            legend pos=south east,
            legend style={font=\small,draw=none},
            grid=both,
            major grid style={dashed,line width=0.5pt,draw=gray!50},
            minor grid style={dotted,line width=0.25pt,draw=gray!25},
            yticklabel style={/pgf/number format/fixed},
            x tick label style={rotate=45,anchor=east},
            width=2.2in, height=2in,
        ]
        \addplot[
            color=red, mark=*, mark options={fill=red},
            error bars/.cd, y dir=both, y explicit, error bar style={line width=0.8pt},
        ] table[x=RTT, y=\ycol, y error=\yerrcol, col sep=comma,each nth point=3] {figures/metrics_AQMn3_Reno.csv};
        \addplot[
            color=blue, mark=square*, mark options={fill=blue},
            error bars/.cd, y dir=both, y explicit, error bar style={line width=0.8pt},
        ] table[x=RTT, y=\ycol, y error=\yerrcol, col sep=comma,each nth point=3] {figures/metrics_noAQM_Reno.csv};
        \end{axis}
    \end{tikzpicture} 
    \caption{Performance evaluation metrics across varying network propagation delays for TCP NewReno variant. The common legend appears in the left plot alone.}
    \label{fig:perf_eval_results_NewReno}
        \vspace{-2mm}
\end{figure*}
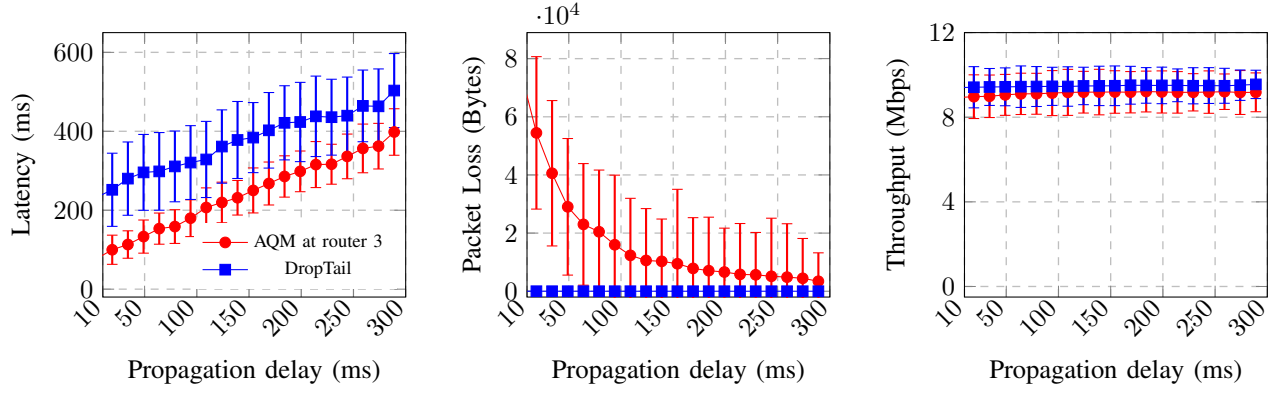


\section{Conclusions}
\label{sec:Conclude}
Despite significant research attention on Active Queue Management (AQM) strategies, adoption and deployment remain sparse. For generic networks, current proposals for AQM deployment dictate that edge routers have them. However, this seems very simplistic given that a bottleneck could occur in the core network despite queues being controlled at the edge. Wide-spread deployment remains infeasible due to the scale and heterogeneity of today's networks. 

We studied AQM deployment in a specific class of networks where routers/switches have a topological hierarchy, form acyclic paths, and adopt multipath routing. This topology is typical of enterprise networks and data centre networks, where network devices have an inherent hierarchy and routing is performed through protocols such as Equal Cost Multipath Routing. Owing to the link between stability and performance in TCP-AQM networks, we approach this problem from a control-theoretic perspective. A closed-loop model was developed for network carrying TCP flows, where one needs to identify a router for AQM deployment. It assumed that the flow dynamics matrix, which dictates the flow of traffic between the routers is known. 
Stability results show that deploying AQM at some routers has a more stabilising effect as compared to others. We then defined a Katz centrality-based metric and showed that it can be used to identify the router for AQM deployment. Stability wise, this metric outperforms the choice of deploying the AQM at the edge alone. Having established the design guidelines for AQM deployment to ensure network stability, we used packet-level simulations to examine if improved stability indeed guarantees better performance. It was established that the proposed guidelines for AQM deployment indeed ensure reduced latency. The proposed solution is particularly suited for Software Defined Network architectures where the a centralized controller performs routing, and hence the traffic flow in the network is known. The Katz centrality computation can be performed by the SDN controller, and the AQM deployment decision can be implemented through Southbound APIs.

\balance
\bibliographystyle{unsrt}  
\bibliography{main}

@String{Computing = "Computing" }

@String{Computer = "{IEEE} Computer" }

@String{Psychometrika = "Psychometrika" }

@String{Springer = "Springer-Verlag" }

@article{manjunath2018stability,
  title="{Stability and performance of Compound TCP with a Proportional Integral queue policy}",
  author={Manjunath, Sreelakshmi and Raina, Gaurav},
  journal={IEEE Transactions on Control Systems Technology},
  volume={27},
  number={5},
  pages={2139--2155},
  year={2019},
}

@Article{floyd1993random,
  title="{Random early detection gateways for congestion avoidance}",
  author={Floyd, Sally and Jacobson, Van},
  journal={IEEE/ACM Transactions on Networking},
  volume={1},
  number={4},
  pages={397--413},
  year={1993},
  publisher={IEEE}
}

@Article{gettys2012bufferbloat,
  title="{Bufferbloat: dark buffers in the internet}",
  author={Gettys, Jim and Nichols, Kathleen},
  journal={Communications of the ACM},
  volume={55},
  number={1},
  pages={57--65},
  year={2012},
  publisher={ACM}
}

@Article{hollot2002analysis,
  title="{Analysis and design of controllers for AQM routers supporting TCP flows}",
  author={Hollot, Christopher V and Misra, Vishal and Towsley, Donald and Gong, Weibo B},
  journal={IEEE Transactions on Automatic Control},
  volume={47},
  number={6},
  pages={945--959},
  year={2002},
  publisher={IEEE}
}

@Article{misra2000fluid,
  title="{Fluid-based analysis of a network of AQM routers supporting TCP flows with an application to RED}",
  author={Misra, Vishal and Gong, Wei-Bo and Towsley, Don},
  journal={ACM SIGCOMM Computer Communication Review},
  volume={30},
  number={4},
  pages={151--160},
  year={2000}
}

@Article{nichols2012controlling,
  title="{Controlling queue delay}",
  author={Nichols, Kathleen and Jacobson, Van},
  journal={Communications of the ACM},
  volume={55},
  number={7},
  pages={42--50},
  year={2012},
  publisher={ACM}
}

@Inproceedings{pan2013pie,
  title="{PIE: A lightweight control scheme to address the bufferbloat problem}",
  author={Pan, Rong and Natarajan, Preethi and Piglione, Chiara and Prabhu, Mythili Suryanarayana and Subramanian, Vijay and Baker, Fred and VerSteeg, Bill},
  booktitle={Proceedings of 14th IEEE International Conference on High Performance Switching and Routing (HPSR)},
  pages={148--155},
  year={2013}
}

@Article{raina2005part,
  title="{Part II: Control theory for buffer sizing}",
  author={Raina, Gaurav and Towsley, Don and Wischik, Damon},
  journal={ACM SIGCOMM Computer Communication Review},
  volume={35},
  number={3},
  pages={79--82},
  year={2005},
  publisher={ACM}
}

@Inproceedings{raina2005buffer,
  title="{Buffer sizes for large multiplexers: TCP queueing theory and instability analysis}",
  author={Raina, Gaurav and Wischik, Damon},
  booktitle={Next Generation Internet Networks},
  pages={173--180},
  year={2005},
  organization={IEEE}
}

@misc{manjunath2019compound,
    title="{Compound TCP with Random Early Detection (RED): stability, bifurcation and performance analyses}",
    author={Sreelakshmi Manjunath and Gaurav Raina},
    year={2019},
    eprint={1907.06302},
    archivePrefix={arXiv},
    primaryClass={cs.NI}
}

@Inproceedings{tan2006compound,
  title="{A compound TCP approach for high-speed and long distance networks}",
  author={Tan, Kun and Song, Jingmin and Zhang, Qian and Sridharan, Murad},
  booktitle={Proceedings of INFOCOM},
  pages={1--12},
  year={2006},
  organization={IEEE}
}

@Article{liu2008tcp,
  title="{TCP-Illinois: A loss-and delay-based congestion control algorithm for high-speed networks}",
  author={Liu, Shao and Ba{\c{s}}ar, Tamer and Srikant, R},
  journal={Performance Evaluation},
  volume={65},
  number={6-7},
  pages={417--440},
  year={2008},
  publisher={Elsevier}
}

@Article{arxiv_version,
  title="{Compound TCP with Random Early Detection (RED): stability, bifurcation and performance analyses}",
  author={Manjunath, Sreelakshmi and Raina, Gaurav},
  journal={arXiv Preprint},
  pages={},
  year={2019},
  publisher={}
}

@Inproceedings{mrozowski2009aqm,
  title     = "{On the deployment of AQM algorithms in the Internet}",
  author    = {Mrozowski, P. and Chydzinski, A.},
  booktitle = {Proceedings of the 11th WSEAS International Conference on Mathematical Methods and Computational Techniques in Electrical Engineering},
  pages     = {276--281},
  year      = {2009},
  organization = {WSEAS}
}

@Article{bideh2016tada,
  title="{TADA: An active measurement tool for automatic detection of AQM}",
  author={Bideh, Minoo Kargar and Petlund, Andreas and Griwodz, Carsten and Ahmed, Iffat and Behjati, Razieh and Brunstr{\"o}m, Anna and Alfredsson, Stefan},
  journal={EAI Endorsed Trans. Self-Adaptive Systems},
  volume={2},
  number={8},
  pages={e5},
  year={2016},
  publisher={Citeseer}
}

@article{hale1977retarded,
  title={Retarded functional differential equations: basic theory},
  author={Hale, Jack K},
  journal={Theory of functional differential equations},
  pages={36--56},
  year={1977},
  publisher={Springer}
}

@misc{harville1998matrix,
  title={Matrix algebra from a statistician's perspective},
  author={Harville, David A},
  year={1998},
  publisher={Taylor \& Francis}
}

@book{aastrom2021feedback,
  title={Feedback systems: an introduction for scientists and engineers},
  author={{\AA}str{\"o}m, Karl Johan and Murray, Richard M},
  year={2021},
  publisher={Princeton university press}
}

@book{bullo2020lectures,
  title={Lectures on network systems},
  author={Bullo, Francesco},
  volume={1},
  number={3},
  year={2020},
  publisher={Kindle Direct Publishing Seattle, DC, USA}
}

@article{katz1953new,
  title={A new status index derived from sociometric analysis},
  author={Katz, Leo},
  journal={Psychometrika},
  volume={18},
  number={1},
  pages={39--43},
  year={1953},
  publisher={Springer}
}

@article{cardwell2017bbr,
  title={{BBR}: Congestion-based congestion control},
  author={Cardwell, Neal and Cheng, Yuchung and Gunn, C Stephen and Yeganeh, Soheil Hassas and Jacobson, Van},
  journal={Communications of the ACM},
  volume={60},
  number={2},
  pages={58--66},
  year={2017},
  publisher={ACM New York, NY, USA}
}

@inproceedings{ferlin2014tackling,
  title={Tackling the challenge of bufferbloat in multi-path transport over heterogeneous wireless networks},
  author={Ferlin-Oliveira, Simone and Dreibholz, Thomas and Alay, {\"O}zg{\"u}},
  booktitle={2014 IEEE 22nd International Symposium of Quality of Service (IWQoS)},
  pages={123--128},
  year={2014},
  organization={IEEE}
}

@misc{McFillin2022,
    author= {Adam McFillin},
    title = {Bufferbloat may be solved, but it’s not over yet},
    url = {https://blog.apnic.net/2020/01/22/bufferbloat-may-be-solved-but-its-not-over-yet/},
    note = {Accessed July 2025},
    publisher = {APNIC (January 22, 2020}
}

@article{allman2012comments,
  title={Comments on bufferbloat},
  author={Allman, Mark},
  journal={ACM SIGCOMM Computer Communication Review},
  volume={43},
  number={1},
  pages={30--37},
  year={2012},
  publisher={ACM New York, NY, USA}
}

@article{cerf2014bufferbloat,
  title={Bufferbloat and other {Internet} challenges},
  author={Cerf, Vinton G},
  journal={IEEE Internet Computing},
  volume={18},
  number={5},
  pages={80--80},
  year={2014},
  publisher={IEEE}
}

@article{harrison2023buffer,
  title={Buffer-bloated router? {How to prevent it and improve performance}},
  author={Harrison, Philippa},
  journal={Communications of the ACM},
  volume={66},
  number={6},
  pages={73--77},
  year={2023},
  publisher={ACM New York, NY, USA}
}

@inproceedings{alwahab2018simulation,
  title={A simulation-based survey of active queue management algorithms},
  author={Alwahab, Dhulfiqar A and Laki, S{\'a}ndor},
  booktitle={Proceedings of the 6th International Conference on Communications and Broadband Networking},
  pages={71--77},
  year={2018}
}

@misc{baker2015ietf,
  series       = {Request for Comments},
  number       = {7567},
  howpublished = {RFC 7567},
  publisher    = {RFC Editor},
  doi          = {10.17487/RFC7567},
  url          = {https://www.rfc-editor.org/info/rfc7567/},
  author       = {F. Baker and G. Fairhurst},
  title        = {{IETF Recommendations Regarding Active Queue Management}},
  pagetotal    = {31},
  year         = {2015},
  month        = jul
}

@book{srikant2004mathematics,
  title={The mathematics of Internet congestion control},
  author={Srikant, Rayadurgam and Srikant, R},
  year={2004},
  publisher={Springer}
}

@misc{floyd2004rfc3782,
  title={Rfc3782: The newreno modification to tcp's fast recovery algorithm},
  author={Floyd, Sally and Henderson, Tom and Gurtov, Andrei},
  year={2004},
  publisher={RFC Editor}
}

@article{toopchinezhad2025machine,
  title={Machine learning approaches for active queue management: A survey, taxonomy, and future directions},
  author={Toopchinezhad, Mohammad Parsa and Ahmadi, Mahmood},
  journal={Computer Networks},
  pages={111174},
  year={2025},
  publisher={Elsevier}
}

@misc{de2023rfc,
  title={RFC 9331: The Explicit Congestion Notification (ECN) Protocol for Low Latency, Low Loss, and Scalable Throughput (L4S)},
  author={De Schepper, Koen},
  year={2023},
  publisher={RFC Editor}
}

@article{zhu2006edge,
  title={Edge-based active queue management},
  author={Zhu, L and Ansari, N and Cheng, G and Xu, K},
  journal={IEE Proceedings-Communications},
  volume={153},
  number={1},
  pages={55--60},
  year={2006},
  publisher={IET}
}

@inproceedings{briscoe2023low,
  title={Low latency, low loss, and scalable throughput (L4S) Internet Service: Architecture},
  author={Briscoe, Bob and Schepper, KD and Bagnulo, Marcelo and White, Greg},
  booktitle={RFC 9330},
  year={2023}
}

@techreport{bensley2017data,
  title={Data center TCP (DCTCP): TCP congestion control for data centers},
  author={Bensley, Stephen and Thaler, Dave and Balasubramanian, Praveen and Eggert, Lars and Judd, Glenn},
  year={2017}
}

\end{document}